\documentclass{article}
\usepackage[margin=1in]{geometry}
\usepackage{epsfig,ulem,amssymb,soul}

\usepackage{hyperref}
\usepackage{graphicx}
\usepackage{epstopdf}
\usepackage{amsfonts,amsmath}
\usepackage{bm}
\usepackage{setspace}
\usepackage{comment}
\usepackage{siunitx}
\usepackage{mathrsfs}
\usepackage{multirow}
\usepackage{subcaption}
\usepackage[hang,flushmargin,bottom]{footmisc}
\usepackage{mathtools}
\usepackage{lipsum}
\usepackage{array}
\usepackage{enumerate}
\usepackage{ifthen}
\usepackage[affil-it]{authblk}
\usepackage{cite}
\usepackage{leftidx}
\usepackage{relsize}
\usepackage{braket}
\usepackage{float}
\usepackage{fontenc}

\usepackage[autostyle]{csquotes}
\usepackage[dvipsnames]{xcolor}

\newcommand{\etal}{\textit{et al}.~}

\title{Fractional-Order Structural Stability: Formulation and Application to the Critical Load of Slender Structures}
\author[1]{Sai Sidhardh}
\author[1]{Sansit Patnaik}
\author[1]{Fabio Semperlotti}
\affil[1]{School of Mechanical Engineering, Ray W. Herrick Laboratories, Purdue University, West Lafayette, IN 47907}
\date{}

\begin{document}

\maketitle
\begin{abstract}
    This study presents the framework to perform stability analysis of nonlocal solids whose response is formulated according to the fractional-order continuum theory. In this formulation, space fractional-order operators are used to capture the nonlocal response of the medium by introducing nonlocal kinematic relations. First, we use the geometrically nonlinear fractional-order kinematic relations within an energy based approach to establish the Lagrange-Dirichlet stability criteria for fractional-order nonlocal structures. This energy based approach to nonlocal structural stability is possible due to a positive-definite and thermodynamically consistent definition of deformation energy enabled by the fractional-order kinematic formulation. Then, the Rayleigh-Ritz coefficient for critical load is derived for linear buckling conditions. The fractional-order formulation is finally used to determine critical buckling loads of slender nonlocal beams and plates using a dedicated fractional-order finite element solver. Results establish that, in contrast to existing studies, the effect of nonlocal interactions is observed on both the material and the geometric stiffness, when using the fractional-order kinematics approach. We support these observations quantitatively with the help of case studies focusing on the critical buckling response of fractional-order nonlocal slender structures, and qualitatively via direct comparison of the fractional-order approach with the classical nonlocal approaches.\\
    
\noindent All correspondence should be addressed to: \textit{ssidhard@purdue.edu} and \textit{fsemperl@purdue.edu}

\end{abstract}

\section{Introduction}

The stability analysis of structures with particular reference to the identification of the critical buckling load is a canonical problem in structural analysis and design. An extensive body of literature is available on this topic in the general area of classical (local) elasticity, which is built upon a point-wise correspondence of the kinematic and material variables via the constitutive relations. Comprehensive reviews of the stability of elastic structures following classical elasticity theories can be found in \cite{timoshenko2009theory,cedolin2010stability}. While this class of so-called local approaches has been, and still is, a fundamental tool to model the behavior of solids, experimental observations have shown that the nonlocal interactions between extended areas of the solid (i.e. between distant points) can have a non-negligible effect on the global response of the medium. These effects, which are a macroscopic manifestation of long-distance interactions between distant points, are not accounted for in classical local theories. Although nonlocal effects have been traditionally restricted to the context of micro- and nano-scale systems \cite{arash2012review,behera2017recent}, examples can be found in a broader range of applications including macro-scale complex media such as sandwich structures as well as functionally graded and porous materials \cite{gurevich1995velocity,romanoff2020review}. 

During the past several decades, several theories have been proposed to model the effect of the nonlocal interactions in elastic solids. Prominent theories were proposed by Kro\"ner \cite{kroner1967elasticity} and Eringen \etal \cite{eringen1972nonlocal} involving strain-based integral constitutive relations. These approaches accounted for the nonlocal interactions within the constitutive relations via a convolution of the local strain with a kernel defined over the domain of influence. In the context of stability analysis, the critical buckling load of slender structures performed using this strain-based integral formulation \cite{tuna2017bending,taghizadeh2016beam} predicted a consistent reduction of the critical loads due to the nonlocal effect. While these strain-based integral formulations were powerful and somewhat very intuitive, the integral definition of the constitutive relation \cite{eringen1972nonlocal} belongs to an ill-posed class of integral equations involving Fredholm integral equations of the first kind, which do not admit unique solutions. Successively, gradient based models of nonlocal elasticity were developed in order to circumvent the issues typical of implicit integral formulations \cite{eringen1983differential}. In most cases, the differential equivalent of the single-phase model \cite{pradhan2009small,reddy2007nonlocal,phadikar2010variational,taghizadeh2016beam} predicted a consistent reduction of the critical loads caused by the nonlocal effect, however paradoxical observations were noted for certain choices of loading and boundary conditions. These observations could be attributed to the non-self adjoint nature of the linear operators obtained following the differential models for Eringen's nonlocal elasticity \cite{challamel2014nonconservativeness,romano2017stress}. Also, note that the differential models are equivalent to their integral counterparts only under certain assumptions for the kernel used in the convolution integral defined assuming an unbounded medium \cite{eringen1972linear}. To address this important issue a two-phase definition (i.e. local/non-local) of the constitutive relations was proposed. This definition admits unique solutions and is generally well-posed in nature leading to self-adjoint linear operators \cite{polizzotto2001nonlocal}. The critical load analysis performed using this two-phase formulation \cite{zhu2017buckling} also predicted a consistent reduction of the critical loads caused by the nonlocal effects. {While the two-phase models present unique solutions with a consistent nonlocal nature across loading and boundary conditions}, {this characteristic property is lost in the limit of vanishing local fractions, as also noted in \cite{romano2017stress}. More specifically, in the limit of a vanishing local fraction, the inherent ill-posedness of fully strain-driven integral problem is not eliminated in the two-phase model. This consideration is at the basis of restrictions on the parameter space that determines the ratio of the local and nonlocal mixture. Further, these strain-driven integral models also do not satisfy the thermodynamic balance laws in a rigorous manner.} 
{More specifically, it has been observed that thermomechanical deformation, obtained via this approach, satisfies the second law of thermodynamics only in a weak (integral) sense and not in a strong (localized) manner \cite{polizzotto2001nonlocal,de2008variational,sidhardh2020thermoelastic}}.  The above observations highlight that there are still some important limitations in the existing nonlocal elasticity theories that affects, although are not limited to, the stability analysis of nonlocal structures.

Recall that, in classical elasticity, the critical load is the ratio of the material and the geometric stiffness of the structures (Rayleigh-Ritz coefficient) \cite{timoshenko2009theory}. In the context of nonlocal elasticity, it is expected that the effects of the nonlocal interactions will be realized upon both of these stiffness terms. However, studies employing the Eringen's strain-integral models of nonlocal elasticity attributed the consistent decrease in the critical load to a reduction of material stiffness caused by nonlocal effects, while the geometric stiffness was essentially unaffected \cite{tuna2017bending,taghizadeh2016beam}. In contrast, the decrease in critical loads predicted by differential models (for those cases not leading to paradoxical observations) was attributed to an increase in geometric stiffness of the structure while the material stiffness was left unaffected \cite{pradhan2009small,taghizadeh2016beam}. A comparison of these contrasting approaches for the calculation of the critical load of nonlocal structures indicates that the strain-integral model predicts a sharper reduction over its equivalent differential model\cite{norouzzadeh2017pre}. Following the above discussion, it is clear that the both the strain-based integral and differential approaches modify \textit{either} the material \textit{or} the geometric stiffness\cite{norouzzadeh2017pre}. However, the physical realization of the nonlocal effects should not be limited to either one of these structural stiffness terms. Clearly there is a gap in proper accounting for the nonlocal effects on the structural stiffness terms, and a clear understanding of this would be required for the stability studies of nonlocal structures.

Recently, the development of fractional-order continuum theories for nonlocal elasticity has offered alternative methodologies that could potentially help filling this gap \cite{carpinteri2014nonlocal,sumelka2014thermoelasticity,patnaik2019generalized}. In recent years, fractional calculus has garnered increasing attention thanks to its many applications in different fields of science and engineering. Successful applications include, to name a few, constitutive modeling of viscoelastic materials to study memory effects \cite{mainardi1996fractional,patnaik2020application}, nonlocal effects across multiple spatial scales \cite{carpinteri2014nonlocal,patnaik2020towards}, dissipation in heat transfer \cite{povstenko2013fractional}. Numerous models for nonlocal elasticity based on the fractional calculus have also been proposed \cite{cottone2009elastic,carpinteri2014nonlocal,sumelka2014thermoelasticity,patnaik2019generalized}. Among the aforementioned studies, studies based on fractional-order kinematic approaches are particularly exciting since they have been able to address key limitations of both integral and gradient based approaches to nonlocal elasticity \cite{patnaik2019generalized,patnaik2020towards}. More specifically, modeling nonlocal interactions at the level of the kinematics in a frame-invariant and dimensionally consistent manner, allows obtaining localized material constitutive relations free from nonlocal residual terms. The resulting nonlocal models allowed the rigorous application of the thermodynamic principles without any physical inconsistency \cite{sidhardh2020thermoelastic}. {In other terms, the fractional-order kinematic approach allows for a strong (or localized) imposition of the first and second laws of thermodynamics at each point within the continuum \cite{sidhardh2020thermoelastic}. This result is unlike the classical nonlocal approaches based on the integral-form of the material constitutive relations that, instead, allow only a weak imposition of the thermodynamic balance laws over the entire domain.}
Further, the positive-definite deformation energy density achieved with this definition guarantees the uniqueness of the solution and allows the application of variational principles as well as the development of finite element based solutions \cite{patnaik2019FEM,patnaik2020plates}. This fractional-order formulation has been employed to study the effects of nonlocal elasticity on both the linear and geometrically nonlinear response of beams and plates \cite{patnaik2019FEM,sidhardh2020geometrically,patnaik2020plates,patnaik2020geometrically}. Large deformation analysis of nonlocal structures can be effectively carried out using geometrically nonlinear fractional-order kinematic relations \cite{sidhardh2020geometrically,patnaik2020geometrically}. This framework facilitated by the fractional-order models provide the foundation required for an energy-based stability analysis of nonlocal structures.

In this study, {building upon the existing geometrically nonlinear fractional-order kinematic approach to nonlocal continuum theory, we develop a framework for the stability analysis of nonlocal slender structures.} As will be shown later, the fractional-order kinematic relations allow the nonlocal effects to be accounted for on both the material and the geometric stiffness terms. The objective of the current work is two-fold. First, the conditions necessary to achieve structural stability of fractional-order nonlocal solids are derived following an energy approach. As part of this goal, we establish the Lagrange-Dirichlet theorem for fractional-order continua and apply it to obtain the critical loads for buckling of nonlocal structures. Note that the energy based approach to stability is possible due to the positive-definite potential energy characteristic of the fractional-order models for nonlocal elasticity \cite{patnaik2019FEM,sidhardh2020thermoelastic}. Second, we apply the stability theory to perform a critical load analysis for the linear buckling of fractional-order beams and plates. For this purpose, {we make use of the fractional finite element model (f-FEM) for numerical solution of the eigenvalue stability problem} and to perform a parametric analysis to assess the effect of the fractional-order nonlocality on the critical buckling load.

The remainder of the paper is structured as follows: we begin with the development of a framework of stability analysis for the fractional-order models of nonlocal elasticity. Later, we use this framework to derive the theoretical and numerical models for nonlocal beams and plates using variational principles. Finally, we use a f-FEM approach to evaluate the critical loads corresponding to the fractional-order nonlocal structures to determine the effect of the nonlocal interactions on the buckling loads.

\section{Constitutive modeling for fractional-order nonlocal elasticity}
\label{sec: constt_model_frac}
In this section, we review the basic constitutive relations for the fractional-order continuum theory\cite{patnaik2019generalized,sidhardh2020geometrically}. We begin with a brief review of the fractional-order kinematic relations and of the constitutive relations for nonlocal solids developed in agreement with thermodynamic principles.

Analogously to classical elasticity models, the fractional-order geometrically nonlinear Lagrangian strain tensor for nonlocal solids is given by \cite{patnaik2019generalized,sidhardh2020geometrically}:
\begin{equation}
    \label{eq: finite_fractional_strain}
\mathop{\textbf{E}}^{\alpha}=\frac{1}{2}\bigr(\nabla^\alpha {\textbf{U}}_X+\nabla^\alpha {\textbf{U}}_X^{T}+\nabla^\alpha {\textbf{U}}_X^{T}\nabla^\alpha {\textbf{U}}_X\bigl)
\end{equation}
where $\textbf{U}(\textbf{X})$ is the Lagrangian displacement field. In the above expression, the fractional-order derivative of the displacement vector with respect to spatial coordinates $\textbf{X} \subseteq \mathbb{R}^3 $ is denoted by $\nabla^\alpha\textbf{U}_X$. The component form for this second order tensor is $ \nabla_{{ij}}^\alpha\textbf{U}_{\textbf{X}} = D^{\alpha}_{X_j}U_i$.
{The space-fractional derivative $D^{\alpha}_\textbf{X}\textbf{U}(\textbf{X})$ is defined using a linear combination of the left- and right-handed Caputo derivatives to the order $\alpha\in(0,1)$ in the following manner \cite{patnaik2019generalized}:
\begin{equation}
\label{eq: RC_definition}
	D^{\alpha}_\textbf{X}\textbf{U}(\textbf{X})=\frac{1}{2}\Gamma(2-\alpha)\big[\textbf{L}_{A}^{\alpha-1}~ {}^C_{\textbf{X}_{A}}D^{\alpha}_{\textbf{X}} \textbf{U}(\textbf{X}) - \textbf{L}_{B}^{\alpha-1}~ {}^C_{\textbf{X}}D^{\alpha}_{\textbf{X}_{B}}\textbf{U}(\textbf{X})\big]
\end{equation}}
where $\Gamma(\cdot)$ is the Gamma function, and ${}^C_{\textbf{X}_{A}}D^{\alpha}_{\textbf{X}}\textbf{U}(\textbf{X})$ and ${}^C_{\textbf{X}}D^{\alpha}_{\textbf{X}_{B}}\textbf{U}(\textbf{X})$ are the left- and right-handed Caputo derivatives of $\textbf{U}(\textbf{X})$, respectively. While the above expression is a form of the Riesz-Caputo derivative defined for {$\alpha\in(0,1)$}, the fractional-order derivative $D^{\alpha}_\textbf{X}\textbf{U}(\textbf{X})$ identically reduces to the first integer-order derivative when $\alpha=1$. We merely note that the above definition is different from the classical Riesz derivative defined in \cite{Samko} using a set of Fourier and inverse Fourier transforms.
The terminals of the RC derivative are defined as $\textbf{X}_A=\textbf{X}-\textbf{L}_A$ and $\textbf{X}_B=\textbf{X}+\textbf{L}_B$. Here, $\textbf{L}_A$ and $\textbf{L}_B$ are length scale parameters associated with the fractional-order model for nonlocal elasticity. The domain enclosed by the terminals $(\textbf{X}_A,\textbf{X}_B)$ defines the horizon of nonlocality at the point $\textbf{X}$. Unlike similar fractional-order continuum theories \cite{sumelka2014thermoelasticity}, in the current formulation, the length scale parameters are considered to be position-dependent. This approach allows an appropriate truncation of the length scales to address asymmetric nonlocal horizon at the physical discontinuities in the domain \cite{patnaik2019generalized}. The parameter $\frac{1}{2}\Gamma(2-\alpha)$ along with the length scales, in the RC definition, ensures the frame invariance of the deformation gradient tensor. Further discussion regarding the objectivity and the physical interpretation of the fractional-order model can be found in \cite{patnaik2019generalized,patnaik2019FEM}.

The differ-integral nature of the fractional-order derivative used above introduces the effect of nonlocal interactions on the elastic response, at the level of kinematics. To illustrate this aspect, the definition of the RC fractional-derivative given in Eq.~\eqref{eq: RC_definition} can be recast as:
\begin{equation}
\label{eq: rc_simpl_form_combined}
    D_{\textbf{X}}^{\alpha} \left[\textbf{U}(\textbf{X})\right]=\int_{\textbf{X}-\textbf{L}_A}^{\textbf{X}+\textbf{L}_B}~ \mathcal{A} (\textbf{X},\bm{\xi},\alpha)~D^1_{\bm{\xi}}\left[\textbf{U}(\bm{\xi})\right]~\mathrm{d}\bm{\xi}
\end{equation}
where the kernel $\mathcal{A} (\textbf{X},\bm{\xi},\alpha)$ is the $\alpha$-order power-law function connecting the point under study $\textbf{X}$ and another point $\bm{\xi}$ within the domain of influence. The above mathematical statement allows the interpretation of the fractional derivative $ D_{\textbf{X}}^{\alpha} \left[\textbf{U}(\textbf{X})\right]$ as a convolution of the respective integer-order derivatives $D^1_{\textbf{X}}\left[\textbf{U}(\textbf{X})\right]$ weighted by the power-law kernel $\mathcal{A} (\textbf{X},\bm{\xi},\alpha)$ over the domain of influence $(\textbf{X}_A,\textbf{X}_B)$. The power-law kernel in the above expressions behaves analogous to the attenuation function used in the classical definition for integer-order nonlocal elasticity\cite{polizzotto2001nonlocal}. Note also that the power-law kernel satisfies the normalization: $\int_{\textbf{X}_A}^{\textbf{X}_B}\mathcal{A}\mathrm{d}\bm{\xi}=1$ for all the points within the solid. This condition allows recovering local response conditions under uniform field distributions\cite{de2008variational}. The position-dependent length scales for nonlocal horizon of influence allows this condition to be satisfied even for points close to the boundary of the solid (see \cite{patnaik2019generalized}). 

The complete nonlinear expressions for the fractional-order Euler-Lagrange strain-displacement relations given in Eq.~\eqref{eq: finite_fractional_strain} can be simplified to obtain the fractional-order analogues of the von-K\'arm\'an strain-displacement relations. For a geometrically nonlinear elastic response assuming large displacement and moderate rotations, but small strains, the fractional-order von-K\'arm\'an strains are \cite{sidhardh2020geometrically,patnaik2020geometrically}:
\begin{equation}
    \label{eq: von_karman_plate}
    \tilde{\epsilon}_{ij}=\underbrace{\frac{1}{2}\left(D^{\alpha}_{X_j}U_i+D^{\alpha}_{X_i}U_j\right)}_{\tilde{e}_{ij}(\textbf{u})}+\underbrace{\frac{1}{2}\left(D^{\alpha}_{X_i}U_3~D^{\alpha}_{X_j}U_3\right)}_{\tilde{q}_{ij}(\textbf{u},\textbf{u})},~~~i,j=1,2
\end{equation}
where $U_3(\textbf{X})$ is the transverse displacement field, and {$U_i$ and $U_j$ ($i,j=\{1,2\}$)} are the in-plane displacement components. The transverse strains (normal and shear) are simply the linearized forms of the respective expressions available from Eq.~\eqref{eq: finite_fractional_strain}. Here, the linear and quadratic components of the von-K\'arm\'an strain are denoted by $\tilde{\textbf{e}}(\textbf{u})$ and $\tilde{\textbf{q}}(\textbf{u},\textbf{u})$, respectively.

Modeling nonlocal interactions via the kinematic relations allows the definition of localized material constitutive relations to be extended to a fractional-order continuum theory {in a thermodynamically consistent manner} \cite{sidhardh2020thermoelastic}. In other terms, the tensor representing the material properties of the fractional nonlocal model maintains the same form as the classical tensor used in local elasticity. For the case of linear elasticity, the localized material constitutive equations provide a one-to-one relation between the fractional-order strain ($\tilde{\bm{\epsilon}}$) and the nonlocal stress ($\tilde{\bm{\sigma}}$) evaluated at a point within the solid. For the general class of hyperelastic solids with a non-dissipative response, a strictly convex functional $\mathcal{U}[\textbf{u}(\textbf{x})]$ referred to as the deformation energy density can be defined. The constitutive relations for the fractional-order nonlocal solid, obtained from the thermodynamic balance laws, may be written as\cite{sidhardh2020thermoelastic}:
\begin{equation}
\label{eq: constt_helm}
    \tilde{\sigma}_{ij}=\frac{\partial \mathcal{U}(\tilde{\bm{\epsilon}})}{\partial \tilde{\epsilon}_{ij}}
\end{equation}
where the deformation energy density for a linear elastic nonlocal solid is:
\begin{equation}
    \label{eq: helmoltz_simp}
    \mathcal{U}(\tilde{\bm{\epsilon}})=\frac{1}{2}\tilde{\sigma}_{ij}(\tilde{\bm{\epsilon}})~\tilde{\epsilon}_{ij}=\frac{1}{2}~\mathcal{C}_{ijkl}~\tilde{\epsilon}_{ij}~\tilde{\epsilon}_{kl}
\end{equation}
$C_{ijkl}$ in the above expression is the positive-definite fourth-order elastic coefficient tensor. {Note that the potential energy is positive-definite and convex in nature for a positive-definite elasticity coefficient tensor}.
The stability of the elastic law (material stability) for linear elastic solids follows from this strict monotonicity and positive-definite elasticity coefficient tensor. It is clear that the conditions for strong ellipticity of the elastic coefficient tensor for the fractional-order nonlocal solid simply follows from analogous results of the classical theory of elasticity. Therefore, the conditions for material stability of the nonlocal solid are also local in nature. Finally, the constitutive relations for the nonlocal stress in an isotropic solid are given as:
\begin{equation}
    \label{eq: constt_isot}
    \tilde{\sigma}_{ij}(\tilde{\bm{\epsilon}})=\lambda\delta_{ij}\tilde{\epsilon}_{kk}+2\mu\tilde{\epsilon}_{ij}
\end{equation}
where $\lambda$ and $\mu$ are the Lam\'e parameters. The conditions for material stability of the isotropic nonlocal solid are $\mu>0$ and $\lambda+(2/3)\mu>0$, similarly to the classical theory of elasticity. {Extending the Drucker's stability postulate for nonlinear constitutive laws, similar comments can be made regarding the stability of nonlinear fractional-order materials \cite{drucker1963postulate}.}

\section{Stability analysis of fractional-order nonlocal solids}
\label{sec: stability}
As noted by Hill\cite{hill1957uniqueness}, for an elastic solid the stability and uniqueness of adjacent equilibrium positions are intimately related. Hill showed that the incremental position is stable if a unique solution can be obtained for the boundary value problem at this point. This observation allows studying the stability of the adjacent equilibrium position by means of a linearized form of the nonlinear governing equations. The increments for adjacent equilibrium positions are characterized by a continuous variation of the control parameter $\Lambda$. This reduces the current analysis to a study of the stability of equilibrium positions for a continuous variation of this control parameter. 

Although kinetic definitions for stability are more general, assuming non-dissipative elastic structures, we conduct the current analysis using the static stability criterion based on energy considerations. This criterion states that:\\

\noindent \textit{Given a displacement field $\textbf{u}\in \mathbb{H}$, where $\mathbb{H}$ is a Hilbert space equipped with the norm $||\textbf{u}||$, we define the potential energy functional $\Pi[\textbf{u},\Lambda]$. The equilibrium point $(\textbf{u}_e,\Lambda_e)$ is considered stable under the following assumptions:
\begin{enumerate}
    \item {The potential energy functional is differentiable up to the second order at $(\textbf{u}_e,\Lambda_e)$.}
    \item {The second variation $\delta^2 \Pi[\textbf{u}_e,\Lambda_e]$ is positive-definite.}
\end{enumerate}}

The above proposition is the analogue of the classical Lagrange-Dirichlet theorem in the framework of a fractional-order model. This extension of the Lagrange-Dirichlet theorem to the fractional-order model is possible due to the thermodynamically consistent definition for internal energy density given in Eq.~\eqref{eq: helmoltz_simp}, for the elastic response of a fractional-order nonlocal solid. In terms of kinetic conditions for stability, this can be interpreted as a bounded response of the nonlocal solid subject to perturbations at $(\textbf{u}_e,\Lambda_e)$. 

The condition described above for static stability translates into the positive-definite nature of the Hessian of the \textit{potential} energy function, referred to as the \textit{tangent} stiffness matrix. As discussed earlier, the deformation energy function for fractional-order nonlocal solid given in Eq.~\eqref{eq: helmoltz_simp} is strictly convex. This ensures that the Hessian of the \textit{deformation} energy density, also referred to as the \textit{elastic} stiffness tensor, is positive-definite. However, it is worth noting that the increase of the control parameter $\Lambda$ can result in a violation of the strong convexity of the \textit{tangent} stiffness tensor. In the following, we study the conditions leading to the onset of instability upon increasing $\Lambda$ and identify specific critical value $\Lambda_c$ for a fractional-order solid. 

The total potential energy $\Pi[\textbf{u},\Lambda]$ of a nonlocal structure occupying a domain $\Omega$ is expressed in terms of the deformation energy density $\mathcal{U}(\tilde{\bm{\epsilon}}(\textbf{u}))$, defined for nonlocal strains in Eq.~\eqref{eq: helmoltz_simp}, and the work done by external surface loads $\textbf{f}(\Lambda)$ applied on the boundary $\partial \Omega^\sigma$. The expression for $\Pi[\textbf{u},\Lambda]$ is:
\begin{equation}
\label{eq: pot_stability}
    \Pi[\textbf{u},\Lambda]=\int_\Omega \mathcal{U}(\tilde{\bm{\epsilon}}(\textbf{u}))~\mathrm{d}V-\int_{\partial \Omega^\sigma} \textbf{f}(\Lambda)\cdot\textbf{u}~\mathrm{d}A
\end{equation}
 The definitions of the deformation energy density in Eq.~\eqref{eq: helmoltz_simp} and of the von-K\'arm\'an strain-displacement relations in Eq.~\eqref{eq: von_karman_plate} may be used in the above expression to arrive at the following result for the first variation of the potential energy:
\begin{equation}
    \delta \Pi =\int_{\Omega} \tilde{\bm{\sigma}}:\left[\tilde{\textbf{e}}(\delta\textbf{u})+\tilde{\textbf{q}}\left(\textbf{u},\delta \textbf{u}\right)\right]~\mathrm{d}V-\int_{\partial \Omega^\sigma} \textbf{f}(\Lambda)\cdot \delta \textbf{u}~\mathrm{d}A=0
\end{equation}
As shown in \cite{patnaik2019FEM}, the solutions to the above equation $(\textbf{u}_e,\Lambda_e)$ serve as the equilibrium points for the static response of a nonlocal solid. According to the Lagrange-Dirichlet theorem, the equilibrium state $(\textbf{u}_e,\Lambda_e)$ is stable if $\delta^2 \Pi[\textbf{u}_e,\Lambda_e]>0$. For this, the second variation of the potential energy $\delta^2 \Pi$ evaluated at the equilibrium point $(\textbf{u}_e,\Lambda_e)$ is given by:
\begin{equation}
\label{eq: second_derivative}
    \delta^2 \Pi =\int_\Omega ~\left[\tilde{\textbf{e}}(\delta\textbf{u})+\tilde{\textbf{q}}\left(\textbf{u}_e,\delta \textbf{u}\right)\right]:\textbf{C}:\left[\tilde{\textbf{e}}(\delta\textbf{u})+\tilde{\textbf{q}}\left(\textbf{u}_e,\delta \textbf{u}\right)\right]+\tilde{\bm{\sigma}}_e:\tilde{\textbf{q}}(\delta \textbf{u},\delta \textbf{u})\mathrm{d}V>0
\end{equation}
where $\tilde{\bm{\sigma}}_e$ is the equilibrium stress evaluated at $(\textbf{u}_e,\Lambda_e)$. 
Following the proposition given above, the critical state may now be identified to be the limit of the stability at which the second variation ceases to be positive definite. Thus, an equilibrium point can be considered the critical point of stability $(\textbf{u}_c,\Lambda_c)$ if:
\begin{equation}
    \int_\Omega \{(\tilde{\textbf{e}}(\delta \textbf{u})+\tilde{\textbf{q}}(\textbf{u}_c,\delta \textbf{u})):\textbf{C}:(\tilde{\textbf{e}}(\delta\textbf{u})+\tilde{\textbf{q}}(\textbf{u}_c,\delta\textbf{u}))+\tilde{\bm{\sigma}}(\Lambda_c): \tilde{\textbf{q}}(\delta \textbf{u},\delta\textbf{u})\}~\mathrm{d}V=0
\end{equation}
{While the above equation may be solved to determine the critical load ($\Lambda_c$) for the \textit{nonlinear} buckling of fractional-order nonlocal solids, in this study, we focus only on \textit{linear} buckling. Analogous to classical approaches, we make certain assumptions to obtain the critical loads for linear buckling \cite{reddy2006theory}. More specifically, we assume a proportional loading force ($\textbf{f}(\Lambda)=\Lambda\textbf{f}^0$, $\textbf{f}^0$ being a representative force vector) and  a small {deformation at the critical point} to reduce the above nonlinear equation to the following linearized form}:
\begin{equation}
    \int_\Omega \{\tilde{\textbf{e}}(\delta \textbf{u}):\textbf{C}:(\tilde{\textbf{e}}(\delta\textbf{u}))+\Lambda_c~\tilde{\bm{\sigma}}^0: \tilde{\textbf{q}}(\delta \textbf{u},\delta\textbf{u})\}~\mathrm{d}V=0
\end{equation}
where $\tilde{\bm{\sigma}}^0$ is the stress generated in the solid by the representative force $\textbf{f}^0$. Finally, the load proportionality factor at the critical point corresponding to the \textit{linear} buckling of nonlocal elastic solids is given by:
\begin{equation}
\label{eq: crit_load}
     \Lambda_c= \text{min}_{\textbf{u}} \left[ -\frac{\int_\Omega  \tilde{\textbf{e}}(\textbf{u}):\textbf{C}:\tilde{\textbf{e}}(\textbf{u})~\mathrm{d}V}{\int_\Omega \tilde{\bm{\sigma}}^0:\tilde{\textbf{q}}(\textbf{u},\textbf{u})~\mathrm{d}V} \right]
\end{equation}
The above expression is the fractional-order analogue of the Rayleigh-Ritz coefficient for critical buckling load used in classical elasticity.
 
The expression of the critical buckling load in Eq.~(\ref{eq: crit_load}) allows for an interesting observations. The numerator of the above expression corresponds to the general stiffness of the structure, while the denominator is referred to as the stability matrix or the geometric stiffness for the nonlocal structure \cite{reddy2014introduction}. {These geometric stiffness terms in the above equation are a result of the geometrically nonlinear strain-displacement relations}. Note that, in obtaining the critical load of the nonlocal solid using the fractional-order kinematic approach, the influence of the nonlocality is realized on both the general stiffness term as well as the geometric stiffness term. Following our discussion in the introduction, we note that this is unlike classical integer-order nonlocal theories that include the nonlocal interactions through material constitutive relations alone. Due to this, the nonlocal interactions modeled by the classical integral theories of nonlocal elasticity affect only the general stiffness while the geometric stiffness terms remain identical to case of local elasticity. The implications of fractional-order kinematics in the geometric stiffness have not been previously noted in the literature. 
 
\section{Buckling of fractional-order slender nonlocal structures}
\label{sec: buckling_of_slender_structures}
In this section, we apply the above formulation to determine the critical loads of nonlocal beams and plates modeled according to the fractional-order continuum theory. {Following the procedure discussed in \S\ref{sec: stability} for a general solid, we begin with geometrically nonlinear strain-displacement relations to derive the system governing equations. This nonlinear framework is required to study critical buckling. Finally, in order to obtain the critical load for linear buckling, we linearize the system equations to setup the eigenvalue problem.}

\subsection{Euler-Bernoulli beams}
\label{subsec: beams}
\begin{figure}[h]
    \centering
    \includegraphics[width=0.6\textwidth]{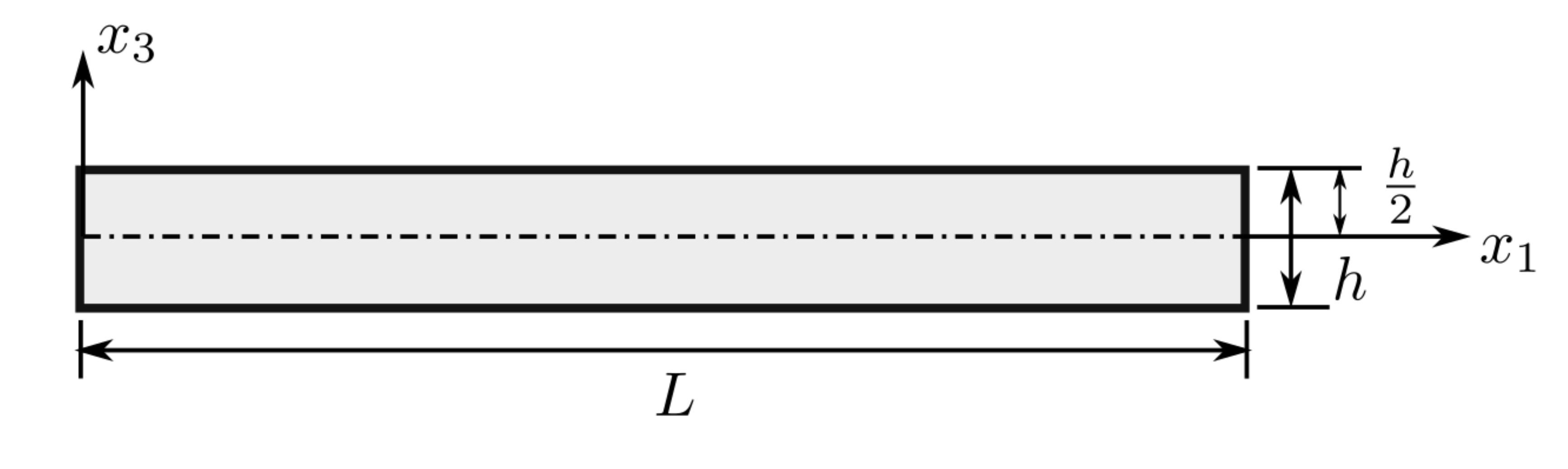}
    \caption{Schematic of the isotropic beam illustrating the Cartesian coordinate axes and a few geometric parameters.}
    \label{fig: beam_schematic}
\end{figure}

In this study, slender beams with geometric length $L$ and height $h$ are chosen such that $L/h>50$. The width of the beam is denoted by $b$. As shown in the schematic in Fig. \ref{fig: beam_schematic}, the Cartesian coordinate axis $x_1$ is aligned along the length of the beam, and the surface $x_3=0$ coincides with the mid-plane. Thus, $x_1=0$ and $x_1=L$ coincide with the ends of the beam, while $x_3=\pm h/2$ are the top and bottom surfaces. 

Under the slender beam assumption, we consider the following Euler-Bernoulli beam displacement theory:
\begin{equation}
    \label{eq: ebt_def}
    u_1(x_1,x_3)=u_0(x_1)-x_3\frac{\partial w_0(x_1)}{\partial x_1}, ~~~u_3(x_1,x_3)=w_0(x_1)
\end{equation}
where, $u_0(x_1)$ and $w_0(x_1)$ are the generalized displacement coordinates defined at a point $X_0(x_1)$ on the reference plane $x_3=0$. They correspond to the axial and transverse displacement fields at $X_0(x_1)$. Hereafter, the functional dependence on the axial coordinate $x_1$ is implied and not mentioned for the sake of brevity. In the following, the fractional-order geometrically nonlinear strains evaluated using the von-K\'arm\'an strain-displacement relations in Eq.~\eqref{eq: von_karman_plate} are:
\begin{equation}
    \label{eq: ebt_Strain}
    \tilde{\epsilon}_{11}=D_{x_1}^\alpha u_0-x_3D_{x_1}^\alpha \left[\frac{\partial w_0}{\partial x_1}\right]+\frac{1}{2}\left(D_{x_1}^\alpha w_0\right)^2
\end{equation}
where $D_{x_1}^\alpha \square$ is a concise notation for the RC fractional-order derivative ${}_{{x_1}-l_A}D^\alpha_{{x_1}+l_B}$ in the $x_1$ direction. Using Eq.~\eqref{eq: constt_isot}, the axial stress in the nonlocal beam is obtained as:
\begin{equation}
    \tilde{\sigma}_{11}(x_1)=E\tilde{\epsilon}_{11}(x_1)
\end{equation}
where $E$ is the Young's modulus for the isotropic solid. Non-zero transverse shear stresses may be neglected under the slender beam assumption. As shown in Eq.~\eqref{eq: pot_stability}, the governing equations of the nonlocal beam subject to external loads may be developed from the principle of minimum potential energy: $\delta \Pi=0$. The resulting nonlinear governing differential equations of equilibrium are\cite{sidhardh2020geometrically}:
\begin{subequations}
\label{eq: beam_governing_equations}
\begin{equation}
\label{eq: beam_axial_gde}
        \mathfrak{D}^{\alpha}_{x_1} {N}_{11}(x_1) + F_1(x_1) = 0 ~~\forall~x_1\in(0,L)
\end{equation}
\begin{equation}
\label{eq: beam_transverse_gde}
        D^1_{x_1}\left[\mathfrak{D}^{\alpha}_{x_1}{M}_{11}(x_1)\right] + \mathfrak{D}^{\alpha}_{x_1} \left[{N}_{11}(x_1)D_{x_1}^{\alpha} \left[w_0\right]\right] + F_3(x_1) = 0 ~~\forall~x_1\in(0,L)
\end{equation}
\end{subequations}
where $F_1(x_1)$ and $F_3(x_1)$ are the distributed forces acting on the nonlocal beam. The essential and natural boundary conditions for the current study are:
\begin{subequations}
\label{eq: beam_all_BCs}
\begin{equation}
\label{eq: beam_axial_bcs}
    {N}_{11}(x_1=L) = {N}_{0} ~~\text{and}~~ \delta u_0(x_1=0) = 0
\end{equation}
\begin{equation}
\label{eq: beam_transverse_moment_bcs}
    {M}_{11}(x_1)=0~~\text{or}~~\delta\left[ D^1_{x_1} w_0\right]=0~~\forall~~x_1\in\{0,L\}
\end{equation}
\begin{equation}
\label{eq: beam_transverse_force_bcs}
    D^1_{x_1} {M}_{11}(x_1) + {N}_{11}(x_1) D^1_{x_1} \left[w_0\right] =0 ~~\text{or}~~\delta w_0=0 ~~\forall~x_1\in\{0,L\}
\end{equation}
\end{subequations}
where $N_0$ is the externally applied surface loads along $x_1$ at the free end. In Eqs.~\eqref{eq: beam_governing_equations} and \eqref{eq: beam_all_BCs}, ${N}_{11}(x_1)$ and ${M}_{11}(x_1)$ are the axial and bending stress resultants in the nonlocal solid. They are defined as follows:
\begin{subequations}
\label{eq: beam_stress_resultants}
\begin{equation}
    {N}_{11}(x_1)=\int_{-b/2}^{b/2}\int_{-h/2}^{h/2}\tilde{\sigma}_{11}(x_1,x_3)~\mathrm{d}x_3~\mathrm{d}x_2
\end{equation}
\begin{equation}
    {M}_{11}(x_1)=\int_{-b/2}^{b/2}\int_{-h/2}^{h/2}x_3~\tilde{\sigma}_{11}(x_1,x_3)~\mathrm{d}x_3~\mathrm{d}x_2
\end{equation}
\end{subequations}

The fractional derivatives $\mathfrak{D}^{\alpha}_{x_1}(\cdot)$ that appear in Eq. \eqref{eq: beam_all_BCs} is the {Riemann Liouville analogue of the RC derivative in Eq.~\eqref{eq: RC_definition}. The Riesz-Riemann Liouville (R-RL) fractional derivative }of order $\alpha$ is defined as:
\begin{equation}
    \label{eq: r_rl_frac_der_def}
    \mathfrak{D}^{\alpha}_{x_1}(\cdot)f(x_1)=\frac{1}{2}\Gamma(2-\alpha) \left[ l_{B}^{\alpha-1} \left({}^{RL}_{x_1-l_{B}} D^{\alpha}_{x_1} f(x_1)\right) - l_{A}^{\alpha-1} \left( {}^{RL}_{x_1}D^{\alpha}_{x_1+l_{A}} f(x_1)\right)\right]
\end{equation}
where $f(x_1)$ is an arbitrary function and ${}^{RL}_{x_1-l_{B}}D^{\alpha}_{x_1}f(x_1)$ and ${}^{RL}_{x_1}D^{\alpha}_{x_1+l_{A}}f(x_1)$ are the left- and right-handed Riemann Liouville derivatives of $f(x_1)$ to the order $\alpha$, respectively. The fractional-order R-RL derivative  $\mathfrak{D}^{\alpha}_{x_1}(\cdot)$ is carried out with respect to the axial coordinate ($x_1$) over the interval $(x_1-l_{B},x_1+l_{A})$. This is unlike the RC fractional derivative $D^{\alpha}_{x}(\cdot)$ defined over the interval $(x_1-l_{A},x_1+l_{B})$. 

The self-adjoint nature of the linear operators in the governing equations follows from the convexity of the deformation energy density used in their derivation. The proof of this property is provided in \cite{patnaik2019FEM}. The positive-definite definition of the deformation energy density given in Eq.~\eqref{eq: helmoltz_simp} and the self-adjoint fractional operators in the governing equations result in a consistent softening of the structure upon inclusion of the nonlocal interactions \cite{patnaik2019FEM,sidhardh2020geometrically}. 
 
For the current study, concerning the identification of the critical load for \textit{linear} buckling, the nonlinear fractional-order governing differential equations given in Eq.~\eqref{eq: beam_all_BCs} are linearized under the assumptions of proportional loading and small deformations at the critical point as discussed in \S\ref{sec: stability}. Considering the case without externally applied distributed loads (i.e. $F_1(x_1)=F_3(x_1)=0$), the linearized governing equations of equilibrium for the Euler-Bernoulli nonlocal beam before the onset of buckling are obtained as:
\begin{subequations}
\label{eq: beam_lin_governing_equations}
\begin{equation}
\label{eq: beam_lin_axial_gde}
        \mathfrak{D}^{\alpha}_{x_1} {N}_{11}(x_1) = 0 ~~\forall~x_1\in(0,L)
\end{equation}
\begin{equation}
\label{eq: beam_lin_transverse_gde}
        D^1_{x_1}\left[\mathfrak{D}^{\alpha}_{x_1}{M}_{11}(x_1)\right] + \overline{{N}}_{0}\mathfrak{D}^{\alpha}_{x_1} \left(D_{x_1}^{\alpha} \left[w_0\right]\right)= 0 ~~\forall~x_1\in(0,L)
\end{equation}
\end{subequations}
where the constant $\overline{N}_{0}$ is the in-plane stress-resultant along $x_1$ at the onset of buckling. In the derivation of the above equations, it is assumed that the beam is straight ($w_0(x_1)=0$) before buckling. Solving the linearized fractional-order governing equation for the axial response given in Eq.~\eqref{eq: beam_lin_axial_gde} and subject to uniform edge loading $N_0$ expressed via natural boundary conditions in Eq.~\eqref{eq: beam_axial_bcs}, we obtain the in-plane stress resultants $\overline{N}_0=N_0$. A detailed derivation of the above results can be extended from similar studies for classical elasticity discussed in \cite{turvey2012buckling}. From the above equations, it is clear that the elastic response in the $x_1$ and $x_3-$directions are decoupled. Therefore, we only proceed with solving the linearized governing equation for the transverse direction given in Eq.~\eqref{eq: beam_lin_transverse_gde}. This is the eigenvalue problem that governs the onset of buckling in a nonlocal beam subject to compressive axial force ${N}_{0}$. The smallest value of the ${N}_{0}$ for which instability sets in is the critical load.

Obtaining analytical solutions to the eigenvalue problem involving fractional-order governing equations is not a trivial task and typically not possible. Thus, we employ a numerical solution based on the f-FEM method developed in \cite{patnaik2019FEM}. Using the method of weighted residuals, the following mathematical expression is equivalent to the governing differential equations in Eq.~\eqref{eq: beam_lin_transverse_gde}:
\begin{equation}
\label{eq: weak_statement_beam}
    \int_0^L \left(D^1_{x_1}\left[\mathfrak{D}^{\alpha}_{x_1}{M}_{11}(x_1)\right] + {{N}}_{0}\mathfrak{D}^{\alpha}_{x_1} \left[D_{x_1}^{\alpha} w_0\right]\right)~ \delta w_0~\mathrm{d}x_1=0 
\end{equation}
where a variation of the transverse displacement field $\delta w_0$ is chosen to be the weight function. Standard integration by-parts gives the following weak-form equivalent of the above governing equation \cite{patnaik2019FEM}:
\begin{equation}
    \label{eq: weak_form_beam}
    \int_{0}^{L}\frac{Eh^3}{12} \left(D_{x_1}^{\alpha} \left[\frac{\partial w_0}{\partial x_1}\right]\right)^2\mathrm{d}x_1-N_0\int_{0}^{L}\left(D_{x_1}^{\alpha}w_0\right)^2\mathrm{d}x_1 = 0
\end{equation}
Using a finite element approximation for the transverse displacement field, the above equation can be reduced to the following algebraic equations:
\begin{subequations}
\label{eq: beam_algebraic}
\begin{equation}
    [K_T^{b}]\{\Delta^b\}=\{0\}
\end{equation}
where the tangent stiffness matrix $[K_T^b]$ is 
\begin{equation}
    [K_T^b]= \underbrace{\int_{0}^{L}\frac{Eh^3}{12} [\tilde{B}_{\mathcal{H},11}]^{T} [\tilde{B}_{\mathcal{H},11}]~\mathrm{d}x_1}_{[K^b]: \text{ Material stiffness}} - N_0 \underbrace{\int_0^L[\tilde{B}_{{\mathcal{H},1}}]^T[\tilde{B}_{\mathcal{H},1}]~\mathrm{d}x_1}_{[G^b]: \text{ Geometric stiffness}} 
\end{equation}
The strain-displacement approximation matrices $[\tilde{B}_{{\mathcal{H},1}}]$ and $[\tilde{B}_{{\mathcal{H},11}}]$ used in the above equation are fractional-order in nature and their expressions follow from the strain-displacement relations in Eq. \eqref{eq: ebt_Strain}. {It is evident from the expression of the geometric stiffness matrix that the modeling of nonlocal response via the fractional-order kinematic approach effects a change in the geometric stiffness of the solid in addition to the material stiffness, as also discussed in \S\ref{sec: stability}. This is unlike classical integer-order approaches to nonlocal elasticity where the geometric stiffness matrix is still local in nature.}
Note that, in the above derivation, we employed the following approximation for the transverse displacement field:
\begin{equation}
    \{w_0(x_1)\}=[\mathcal{H}(x_1)]\{\Delta^b_e(x_1)\}
\end{equation}
where $[\mathcal{H}(x_1)]$ are the one-dimensional $C^1$ ($\mathcal{H}$ermite) approximation functions, and:
\begin{equation}
    \{\Delta^b_e\}^T=\left[w_0^i~~~~\frac{d w_0}{d x_1}^i\right]\bigg\vert_{i=1}^{N_e}
\end{equation}
is the element nodal vector of the generalized displacement coordinates for the $N_e-$noded element.
\end{subequations}

\subsection{Kirchhoff plates}
\label{subsec: plates}
\begin{figure}[h]
    \centering
    \includegraphics[width=0.5\textwidth]{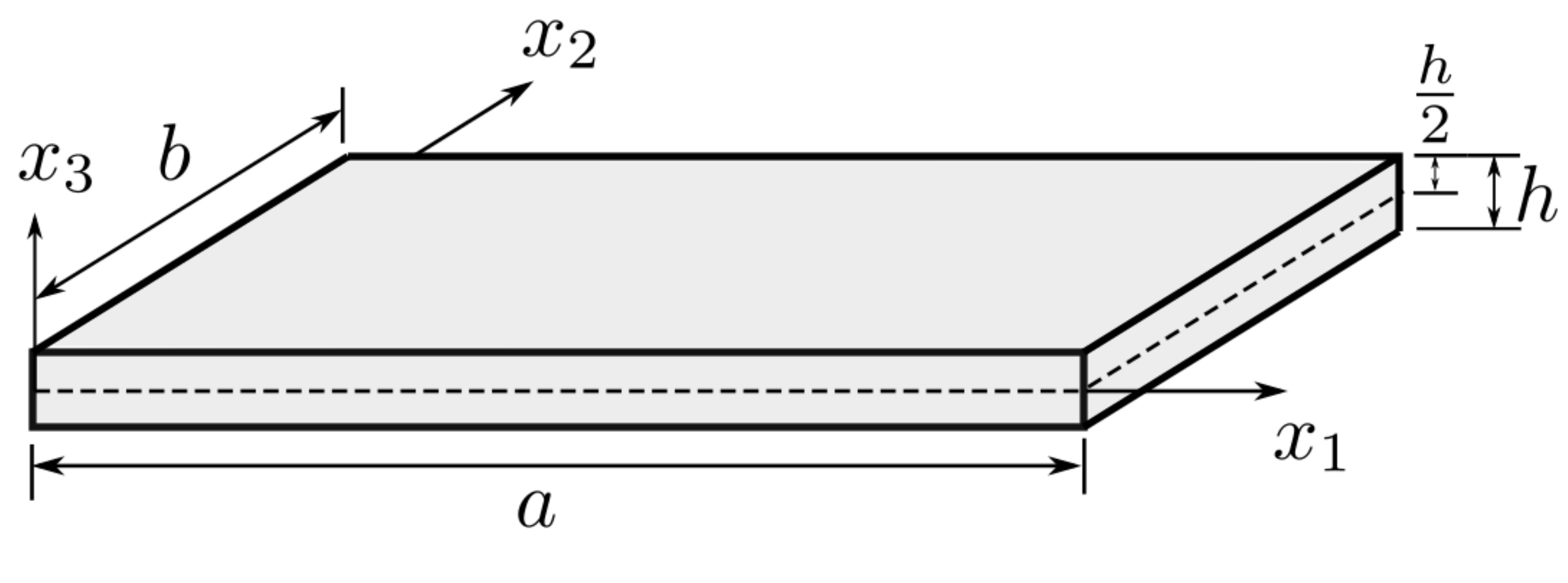}
    \caption{Schematic of the isotropic plate illustrating the Cartesian coordinate axes and relevant geometric parameters.}
    \label{fig: plate_schematic}
\end{figure}

The methodology outlined above is extended to a thin plate with in-plane dimensions $a\times b$ and the thickness $h<a/50$. As shown in the schematic in Fig. \ref{fig: plate_schematic}, the Cartesian coordinates are chosen such that $x_3=0$ is the mid-plane for the plate, and $x_1=0,a$ and $x_2=0,b$ coincide with the transverse free faces of the plate. We consider the displacement field distribution according to the Kirchhoff plate theory:
\begin{equation}
    \label{eq: kirch_def}
    u_1(\textbf{X})=u_0(\textbf{X}_0)-x_3\frac{\partial w_0(\textbf{X}_0)}{\partial x_1}, ~~~u_2(\textbf{X})=v_0(\textbf{X}_0)-x_3\frac{\partial w_0(\textbf{X}_0)}{\partial x_2}, ~~~u_3(\textbf{X})=w_0(\textbf{X}_0)
\end{equation}
where $u_0(\textbf{X}_0)$, $v_0(\textbf{X}_0)$, and $w_0(\textbf{X}_0)$ are the generalized displacement coordinates evaluated at a point $\textbf{X}_0(x_1,x_2)$ on the reference plane $x_3=0$. 
The expressions for the fractional-order geometrically nonlinear strains evaluated following the von-K\'arm\'an strain-displacement relations given in Eq.~\eqref{eq: von_karman_plate} are:
\begin{subequations}
\begin{equation}
    \label{eq: kirch_Strain11}
    \tilde{\epsilon}_{11}=D_{x_1}^\alpha u_0-x_3D_{x_1}^\alpha \left[\frac{\partial w_0}{\partial x_1}\right]+\frac{1}{2}\left(D_{x_1}^\alpha w_0\right)^2
\end{equation}
\begin{equation}
    \label{eq: kirch_Strain22}
    \tilde{\epsilon}_{22}=D_{x_2}^\alpha v_0-x_3D_{x_2}^\alpha \left[\frac{\partial w_0}{\partial x_2}\right]+\frac{1}{2}\left(D_{x_2}^\alpha w_0\right)^2
\end{equation}
\begin{equation}
    \label{eq: kirch_Strain12}
    \tilde{\gamma}_{12}=2\tilde{\epsilon}_{12}=\left(D_{x_1}^\alpha v_0+D_{x_2}^\alpha u_0\right)-x_3\left(D_{x_1}^\alpha \left[\frac{\partial w_0}{\partial x_1}\right]+D_{x_2}^\alpha \left[\frac{\partial w_0}{\partial x_2}\right]\right)+\left(D_{x_1}^\alpha w_0 D_{x_2}^\alpha w_0\right)
\end{equation}
\end{subequations}
Here, $D_{x_1}^\alpha \equiv {}_{{x_1}-l_{A_1}}D^\alpha_{{x_1}+l_{B_1}}$ and $D_{x_2}^\alpha \equiv {}_{{x_2}-l_{A_2}}D^\alpha_{{x_2}+l_{B_2}}$ denote the RC fractional derivatives along $x_1$ and $x_2$. The domains $(x_1-l_{A_1},x_1+l_{B_1})$ and $(x_2-l_{A_2},x_2+l_{B_2})$ provide the horizon of influence for the point $\textbf{X}_0(x_1,x_2)$ along the $x_1$ and $x_2-$directions. The length scales $l_{A_{i}}$ and $l_{B_{i}}$ ($i=1,2$) are the nonlocal length scales in $x_i-$direction.
Using Eq.~\eqref{eq: constt_isot}, the nonlocal stresses in the isotropic plate are obtained as:
\begin{subequations}
    \label{eq: constt_plate}
\begin{equation}
    \tilde{\sigma}_{11}=\frac{E}{1-\nu^2}\left(\tilde{\epsilon}_{11}+\nu\tilde{\epsilon}_{22}\right)
\end{equation}
\begin{equation}
    \tilde{\sigma}_{22}=\frac{E}{1-\nu^2}\left(\nu\tilde{\epsilon}_{11}+\tilde{\epsilon}_{22}\right)
\end{equation}
\begin{equation}
    \tilde{\sigma}_{12}=\frac{E}{2(1+\nu)}\tilde{\gamma}_{12}
\end{equation}
\end{subequations}
where $E$ and $\nu$ are the Young's modulus and Poisson's ratio for the isotropic solid. 

The governing equations of the nonlocal plate based on fractional-order kinematics are derived using the principle of minimum potential energy principle. They are given as \cite{patnaik2020geometrically}:
\begin{subequations}
\label{eq: Kirchoff_GDE}
\begin{equation}
\mathfrak{D}^\alpha_{x_1} {N}_{11} + \mathfrak{D}^\alpha_{x_2} {N}_{12}+F_1 = 0
\end{equation}
\begin{equation}
\mathfrak{D}^\alpha_{x_1} {N}_{12} + \mathfrak{D}^\alpha_{x_2} {N}_{22}+F_2 = 0
\end{equation}
\begin{equation}
\begin{split}
D^{1}_{x_1}\left[\mathfrak{D}^\alpha_{x_1} M_{11}+\mathfrak{D}^\alpha_{x_2} M_{12} \right] + D^1_{x_2} \left[\mathfrak{D}^\alpha_{x_1} M_{12} +\mathfrak{D}^\alpha_{x_2} M_{22}\right]+\mathfrak{D}^\alpha_{x_1} ({N}_{11} D^\alpha_{x_1} w_0 + {N}_{12} D^\alpha_{x_2} w_0 ) \\ +~\mathfrak{D}^\alpha_{x_2} ({N}_{12} D^\alpha_{x_1} w_0 + {N}_{22} D^\alpha_{x_2} w_0 )+F_3 = 0
\end{split}
\end{equation}
\end{subequations}
where $F_i$ ($i=\{1,2,3\}$) are externally applied distributed forces, and ${N}_{\square}$ and ${M}_{\square}$ are the membrane and bending stress resultants evaluated at $\textbf{X}_0$, respectively. They are evaluated using the nonlocal stresses given in Eq.~\eqref{eq: constt_plate} as follows:
\begin{equation}
\label{eq: stress_resultants}
    \{{N}_{11},~{N}_{22},~{N}_{12}\}=\int_{-h/2}^{h/2}\{\tilde{\sigma}_{11},~\tilde{\sigma}_{22},~\tilde{\sigma}_{12}\}\mathrm{d}x_3,~~~~\{{M}_{11},~{M}_{22},~{M}_{12}\}=\int_{-h/2}^{h/2}x_3\{\tilde{\sigma}_{11},~\tilde{\sigma}_{22},~\tilde{\sigma}_{12}\}\mathrm{d}x_3
\end{equation}
The boundary conditions necessary to solve the governing equations given above are:
\begin{subequations}
\label{eq: Kirchoff_BC}
\begin{equation}
\forall x_2 \mid x_1=\{0,a\}:
\left\{
\begin{matrix*}[l]
    \delta u_0=0 & \text{or} & {N}_{11}=N_1\\
    \delta v_0=0 & \text{or} & {N}_{12}=0\\
    \delta w_0=0 & \text{or} & {D}^1_{x_1} {M}_{11} + 2{D}^1_{x_2} {M}_{12} + N_{11}D^1_{x_1} w_0 + {N}_{12} D^1_{x_2} w_0 = 0\\
    \delta D^1_{x_1} w_0 = 0 & \text{or} & {M}_{11}=0
\end{matrix*} \right.
\end{equation}
\begin{equation}
\forall x_1 \mid x_2=\{0,b\}:
\left\{
\begin{matrix*}[l]
    \delta u_0=0 & \text{or} & {N}_{12}=0\\
    \delta v_0=0 & \text{or} & {N}_{22}=N_2\\
    \delta w_0=0 & \text{or} & {D}^1_{x_2} {M}_{22} + 2{D}^1_{x_1} {M}_{12} +  {N}_{12} D^1_{x_1} w_0 + {N}_{22} D^1_{x_2} w_0= 0\\
    \delta D^1_{x_2} w_0 = 0 & \text{or} & {M}_{22}=0
\end{matrix*} \right.
\end{equation}
\end{subequations}
where $N_1$ and $N_2$ are the externally applied uniform surface loads at the free ends in the $x_1$ and $x_2-$directions, respectively. In the above equations, the terms $\mathfrak{D}^\alpha_{x_1}(\cdot)$ and $\mathfrak{D}^\alpha_{x_2}(\cdot)$ are the R-RL fractional-order derivative with respect to $x_1$ and $x_2$. The expressions for these derivatives follow from the definition for R-RL fractional derivative given in Eq.~\eqref{eq: r_rl_frac_der_def}. 

In order to determine the critical load for linear buckling, the above given fractional-order nonlinear governing equations of equilibrium are linearized following the methodology outlined in \S\ref{sec: stability} for general solids and employed for analysis of beams in \S \ref{subsec: beams}. The linearized governing equations for the Kirchhoff plates before the onset of buckling, assuming externally applied distributed loads to be absent, are: 
\begin{subequations}
\label{eq: lin_plate_gde}
\begin{equation}
\mathfrak{D}^\alpha_{x_1} {N}_{11} + \mathfrak{D}^\alpha_{x_2} {N}_{12} = 0
\end{equation}
\begin{equation}
\mathfrak{D}^\alpha_{x_1} {N}_{12} + \mathfrak{D}^\alpha_{x_2} {N}_{22} = 0
\end{equation}
\begin{equation}
\label{eq: lin_Kirchoff_GDE}
\begin{split}
D^{1}_{x_1}\left[\mathfrak{D}^\alpha_{x_1} M_{11}+\mathfrak{D}^\alpha_{x_2} M_{12} \right] + D^1_{x_2} \left[\mathfrak{D}^\alpha_{x_1} M_{12} +\mathfrak{D}^\alpha_{x_2} M_{22}\right]+\overline{N}_{11}  \mathfrak{D}^\alpha_{x_1} [D^\alpha_{x_1} w_0]+\overline{N}_{12}  \mathfrak{D}^\alpha_{x_1} [D^\alpha_{x_2} w_0] \\ +~\overline{N}_{12}  \mathfrak{D}^\alpha_{x_2} [D^\alpha_{x_1} w_0]+\overline{N}_{22}  \mathfrak{D}^\alpha_{x_2} [D^\alpha_{x_2} w_0]=0
\end{split}
\end{equation}
\end{subequations}
where $\overline{N}_{\square}$ are the in-plane stress-resultants before the onset of buckling. We assume two separate cases: (1) uniaxial compression where the plate is subject to externally applied distributed surface loads $N_{1}$ on the transverse faces at $x_1=0$ and $x_1=a$; (2) biaxial compression where the plate is subject to $N_1$ on faces $x_1=0,a$ and $N_2$ on faces $x_2=0,b$. In the absence of shear loads, the linearized fractional-order governing equation for the transverse displacement of the nonlocal plate is given by:
\begin{equation}
\label{eq: lin_Kirchoff_GDE_2}
    \begin{split}
D^{1}_{x_1}\left[\mathfrak{D}^\alpha_{x_1} M_{11}+\mathfrak{D}^\alpha_{x_2} M_{12} \right] + D^1_{x_2} \left[\mathfrak{D}^\alpha_{x_1} M_{12} +\mathfrak{D}^\alpha_{x_2} M_{22}\right]+N_1  \mathfrak{D}^\alpha_{x_1} [D^\alpha_{x_1} w_0]+N_2  \mathfrak{D}^\alpha_{x_2} [D^\alpha_{x_2} w_0]=0
\end{split}
\end{equation}
The above equation corresponds to the biaxial compression, and may be reduced to uniaxial compression by setting, as an example, $N_2=0$. Note that the in-plane stress resultants $\overline{N}_{11}$ and $\overline{N}_{22}$ in Eq. \eqref{eq: lin_Kirchoff_GDE} are equal to the magnitude of the uniform edge loads $N_1$ and $N_2$, respectively. This follows from solving the in-plane governing equations given in Eq. \eqref{eq: lin_plate_gde} subject to boundary conditions given in Eq. \eqref{eq: Kirchoff_BC}. 

As in the case of fractional-order beams, we use the method of weighted residuals to express the following mathematical statement as equivalent to the fractional-order governing equation in Eq.~\eqref{eq: lin_Kirchoff_GDE_2}:
\begin{equation}
\begin{split}
    \int_{0}^{a}\int_{0}^{b} &\left(D^{1}_{x_1}\left[\mathfrak{D}^\alpha_{x_1} M_{11}+\mathfrak{D}^\alpha_{x_2} M_{12} \right] + D^1_{x_2} \left[\mathfrak{D}^\alpha_{x_1} M_{12} +\mathfrak{D}^\alpha_{x_2} M_{22}\right]\right.\\
    &~~~~~~\left.+N_1  \mathfrak{D}^\alpha_{x_1} [D^\alpha_{x_1} w_0]+N_2  \mathfrak{D}^\alpha_{x_2} [D^\alpha_{x_2} w_0]\right) ~\delta w_0~ \mathrm{d}x_2~\mathrm{d}x_1=0
\end{split}
\end{equation}
Integral operations to reduce the above statement into the weak equation for Eq.~\eqref{eq: lin_Kirchoff_GDE_2}, and finite element approximations for the displacement field variables gives the following algebraic equations of equilibrium:
\begin{subequations}
\label{eq: platealgebraic}
\begin{equation}
    [K_T^p]\{\Delta^p\}=\{0\}
\end{equation}
where the tangent stiffness matrix $[K_T^p]$ is given as
\begin{equation}
    [K_T^p]=[K^p]-{N}_{1}[G^p_{1}]-N_2[G^p_2]
\end{equation}
\begin{equation}
\label{eq: stiff_plate}
\begin{split}
    [K^p]=\int_{0}^{a}\int_0^b&\frac{Eh^3}{12(1-\nu^2)} \left([\tilde{B}_{\mathcal{H},11}]^{T} ([\tilde{B}_{\mathcal{H},11}]+\nu~[\tilde{B}_{\mathcal{H},22}])+[\tilde{B}_{\mathcal{H},22}]^{T} ([\tilde{B}_{\mathcal{H},22}]+\nu~[\tilde{B}_{\mathcal{H},11}])\right)\\
    &+\frac{Eh^3}{24(1+\nu)}\left([\tilde{B}_{\mathcal{H},12}]+[\tilde{B}_{\mathcal{H},21}]\right)^T\left([\tilde{B}_{\mathcal{H},12}]+[\tilde{B}_{\mathcal{H},21}]\right)~\mathrm{d}x_1~\mathrm{d}x_2\\
\end{split}
\end{equation}
is the bending stiffness matrix for the fractional-order Kirchhoff plate, and
\begin{equation}
\label{eq: geo_Stiff_plate}
    [G^p_1]=\int_0^a\int_0^b[\tilde{B}_{\mathcal{H},1}]^T[\tilde{B}_{\mathcal{H},1}]~\mathrm{d}x_1\mathrm{d}x_2~~~[G^p_2]=\int_0^a\int_0^b[\tilde{B}_{\mathcal{H},2}]^T[\tilde{B}_{\mathcal{H},2}]~\mathrm{d}x_1\mathrm{d}x_2
\end{equation}
are the geometric stiffness matrix for compression along $x_1$ and $x_2$ directions, respectively. {Note the effect of the fractional-order nonlocality on the geometric stiffness matrices via the fractional-order strain-displacement approximation matrices $[\tilde{B}_{\mathcal{H},\square}]$}. In the derivation of the algebraic equations of equilibrium, we employed the following approximation for the transverse displacement field:
\begin{equation}
        \{w_0(x_1,x_2)\}=[\mathcal{H}(x_1,x_2)]\{\Delta^p_e(x_1,x_2)\}
    \end{equation}
   where $[\mathcal{H}(x_1,x_2)]$ are the two-dimensional $C^1$ ($\mathcal{H}$ermite) approximation functions. The element nodal vector is expressed as:
\begin{equation}
    \{\Delta^p_e\}^T=\left[w_0^i~~~~\frac{d w_0}{d x_1}^i~~~~\frac{d w_0}{d x_2}^i~~~~\frac{d^2 w_0}{d x_1dx_2}^i\right]\bigg\vert_{i=1}^{N_e}
\end{equation}
\end{subequations}
and includes the generalized displacement coordinates for the $N_e-$noded element.
 
\section{Results and discussion}
\begin{figure}[h]
    \centering
    \includegraphics[width=0.7\textwidth]{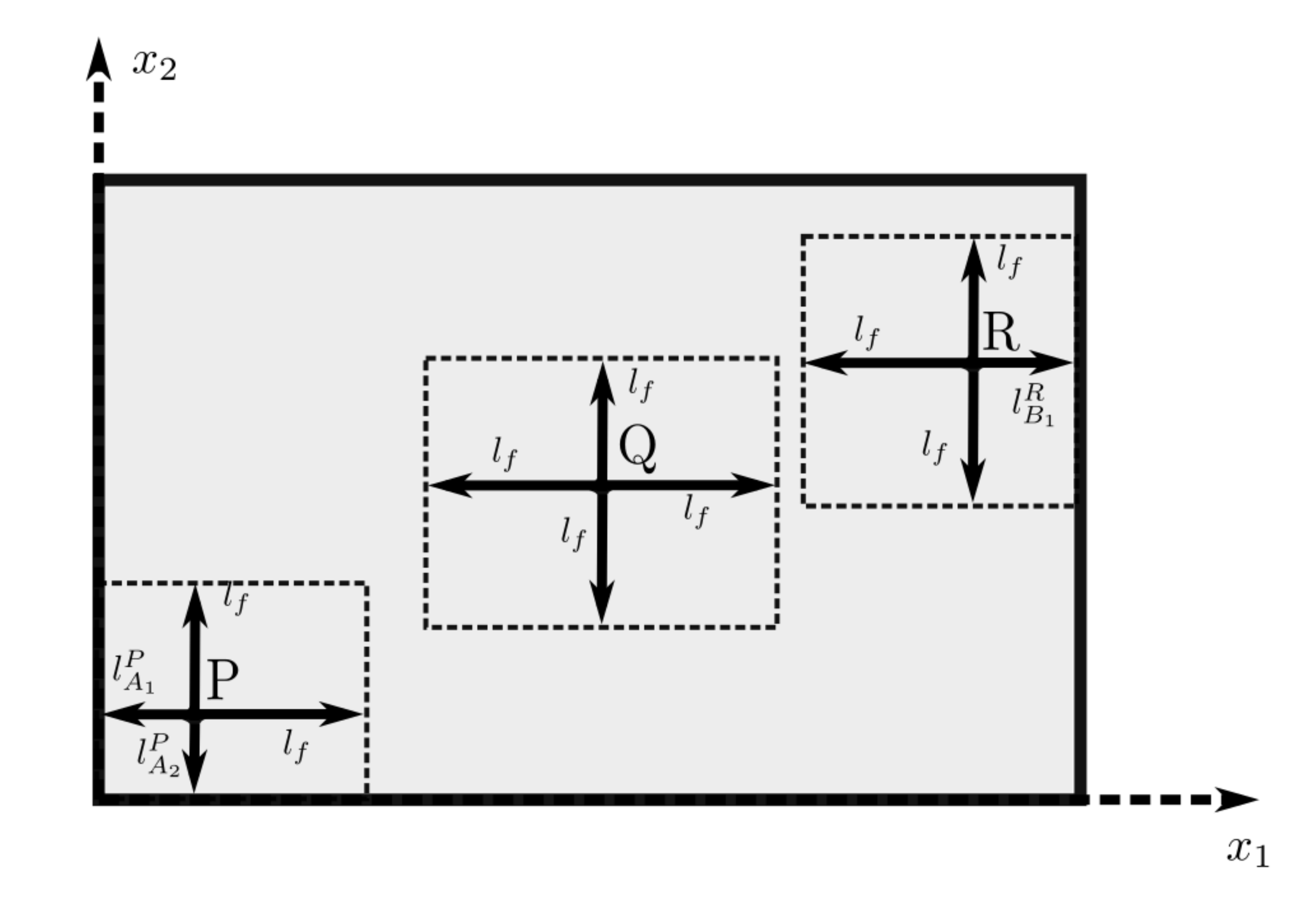}
    \caption{Illustration of position dependent length scales for three different points (P,Q, and R) in the isotropic domain. Note the asymmetry in length scales for points P and R which are close to the boundaries.}
    \label{fig: length_schematic}
\end{figure}

This section reports the results of different types of numerical simulations targeted to provide a quantitative assessment of the effect of fractional-order nonlocality on the critical load of slender structures. More specifically, the f-FEM models for the Euler-Bernoulli beam in Eq.~\eqref{eq: beam_algebraic} and the Kirchhoff plate in Eq.~\eqref{eq: platealgebraic} are used to analyze the effect of the fractional-order and the length scales on the critical load. In all simulations, we assume an isotropic material having the Young's modulus $E=30$ MPa and the Poisson's ratio $\nu=0.3$. 
The nonlocal horizon is assumed to be symmetric for all those points that are sufficiently far from the boundaries. In other terms, the length scales on both sides of the point of interest are assumed to be equal. For a beam this condition translates to $l_A=l_B=l_f$, while for a plate it means $l_{A_1}=l_{B_1}=l_{A_2}=l_{B_2}=l_f$. Clearly, the symmetry of the nonlocal horizons is broken when considering points whose distance from the boundary (in a given direction) is smaller than $l_f$. In this latter case, appropriate truncation of the length scales is performed as indicated in the schematic of Fig. \ref{fig: length_schematic}.

\subsection{Beams}
\label{ssec: beam_results}
In line with the assumptions for the Euler-Bernoulli beam displacement theory, the aspect ratio of the beam was chosen to be $L/h=100$ for the current study. The width of the beam was chosen as $2\times h$. The critical loads reported here were non-dimensionalized as follows\cite{reddy2003mechanics}:
\begin{equation}
    \label{eq: nondim_beam}
    \overline{N}_0=N_0 \times \frac{L^2}{\pi^2 EI}
\end{equation} 
Before presenting the results for the critical loads, we make an important remark on the convergence of the f-FEM. As discussed in \cite{norouzzadeh2017finite}, convergence of the 1D f-FEM depends on the '\textit{dynamic rate of convergence}'. This parameter, which controls the accuracy of the numerical approximation of the convolution integral corresponding to the nonlocal interactions, is defined as $\mathcal{N}^{inf}=l_f/l_e$, where $l_e$ is the size of the uniform FE mesh. The convergence of the f-FEM used in this study is analysed for a clamped-clamped case for different values of the fractional-order parameters and the results are reported in Table \ref{tab: convergence_comp_beam}. The study illustrates the excellent convergence of the normalized eigenvalues that achieve differences of $<$1\% between successive refinements of the FE mesh. Following these results, we used $N^{inf}=24$ for all the simulations presented in this section.

\begin{table}[t!]
    \centering
    \begin{tabular}{c | c |c c c c}
    \hline\hline
       \multirow{2}{14em}{ Fractional Horizon Length, $l_f$} & \multirow{2}{6em}{~~~~~~$N^{inf}$} & \multicolumn{4}{c}{fractional-order, $\alpha$}\\
         \cline{3-6}
         &  & $\alpha=1.0$& $\alpha=0.9$& $\alpha=0.8$ & $\alpha=0.7$ \\
         \hline\hline
         \multirow{4}{5em}{$l_f=0.2L$} & 12 & 4 & 3.9574 & 3.9374 & 3.9200\\
         & 18 & 4 & 3.9529 & 3.9291 & 3.9072\\
         & 24 & 4 & 3.9514 & 3.9251 & 3.8979\\
         & 30 & 4 & 3.9506 & 3.9224 & 3.8908\\
         \hline
          \multirow{4}{5em}{$l_f=0.6L$} & 12 & 4 & 3.7216 & 3.3789 & 2.8594\\
          & 24 & 4 & 3.7238 & 3.3840 & 2.8777\\
          & 36 & 4 & 3.7246 & 3.3871 & 2.8817\\
          & 48 & 4 & 3.7250 & 3.3882 & 2.8817\\
          \hline
          \multirow{4}{5em}{$l_f=L$} & 12 & 4 & 3.8460  & 3.6969 & 3.4806\\
          & 24 & 4 & 3.8484 & 3.7116 & 3.5216\\
          & 36 & 4 & 3.8497 & 3.7182 & 3.5389\\
          & 30 & 4 & 3.8504 & 3.7215 & 3.5466\\
          \hline\hline
    \end{tabular}
    \caption{1D f-FEM mesh convergence study. Non-dimensional critical load of a clamped-clamped beam for different values of the fractional parameters. In all cases, $N^{inf}=24$ guarantees a difference between successive refinements within 1\%.}
    \label{tab: convergence_comp_beam}
\end{table}

The critical loads of fractional-order beams subject to doubly clamped (CC) and simply supported (SS) boundary conditions at $x_1=0,L$ are tabulated in Table \ref{tab: beam} for various fractional-order parameters ($\alpha$ and $l_f$). A non-monotonic variation of the critical load with increasing degree of nonlocality, obtained by reducing $\alpha$ and/or increasing $l_f$, is clearly evident from the table. This observation is unlike the monotonous reduction in critical load noted from similar studies based on classical integer-order nonlocal theories\cite{zhu2017buckling,tuna2017bending,taghizadeh2016beam,norouzzadeh2017pre}. This difference is a direct result of the effect of the nonlocal response on the geometric stiffness. {This aspect is discussed in further detail in the following.} 

\begin{table}[h]
    \centering
    \begin{tabular}{c |c c c c| c c c c}
    \hline\hline
    & \multicolumn{4}{c|}{SS}& \multicolumn{4}{c}{CC}\\
    \hline
      $\alpha$& 1.0 & 0.9 & 0.8 & 0.7& 1.0 & 0.9 & 0.8 & 0.7\\
    \hline
       $l_f/L=$ 0.2 & 1.000 & 0.996 & 0.972 & 0.917 & 4.000 & 3.953 & 3.929 & 3.907\\
       $l_f/L=$ 0.4 & 1.000 & 1.017 & 0.998 & 0.927 & 4.000 & 3.745 & 3.469 & 3.120\\
       $l_f/L=$ 0.6 & 1.000 & 1.069 & 1.095 & 1.046 & 4.000 & 3.720 & 3.376 & 2.877\\
       $l_f/L=$ 0.8 & 1.000 & 1.118 & 1.200 & 1.208 & 4.000 & 3.815 & 3.598 & 3.242\\
       $l_f/L=$ 1.0 & 1.000 & 1.140 & 1.255 & 1.319 & 4.000 & 3.848 & 3.712 & 3.522\\
        \hline\hline
    \end{tabular}
    \caption{Non-dimensional critical loads for a beam subject to various boundary conditions. Results are presented for different values of the fractional constitutive parameters. Recall that SS stands for simply supported and CC for doubly clamped.}
    \label{tab: beam}
\end{table}

As discussed in \S \ref{sec: stability}, contrary to the classical approaches to nonlocal elasticity, adopting the fractional-order kinematic approach to determine the critical load of a nonlocal solid affects both the material and geometric stiffness terms. With the increasing degree of nonlocality, both these stiffness terms are reduced. The decrease in material stiffness (often referred to as softening effect) as a result of the fractional-order kinematics was already documented in \cite{patnaik2019FEM,sidhardh2020geometrically}, and is in agreement with similar observations from classical strain-driven integral approaches to nonlocal elasticity. The reduction of the geometric stiffness due to fractional-order kinematics was not reported in either the fractional- or integer-order studies. {We merely note that, with regards to the fractional-order studies \cite{patnaik2019FEM,sidhardh2020geometrically,patnaik2020plates,patnaik2020geometrically}, the effect of the nonlocality on the geometric stiffness was not reported simply because of the nature of the problems (static and free vibration response) treated in the same studies. Recall however that, the constitutive modeling of the slender structures in this study and the other fractional-order studies in \cite{patnaik2019FEM,sidhardh2020geometrically,patnaik2020plates,patnaik2020geometrically} is still the same.} In this regard, as evident from Eq.~\eqref{eq: crit_load}, the reducing material stiffness (numerator) would result in decreasing the critical load, and the reducing geometric stiffness (denominator) would result in increasing critical load. It follows that the effect of an increasing degree of nonlocality on the critical load is non-monotonic due to an interplay between these contrasting effects, resulting from a simultaneous decrease in the material and geometric stiffness.

To better illustrate the above aspects, we present a numerical study that isolates the effects of nonlocal interactions, modeled by the fractional-derivatives, on the material and geometric stiffness of the nonlocal solid. In an attempt to isolate the effects of the nonlocality produced by the fractional-order kinematics on both stiffness terms, we artificially replace the fractional-derivatives present in their definitions with integer-order derivatives. This process is carried out individually for either the material or the geometric stiffness matrices. For this purpose, the different stiffness matrices in the algebraic equations of the eigenvalue problem developed in Eq. \eqref{eq: beam_algebraic} are modified for the two different parametric studies in the following manner:
\begin{subequations}
\label{eq: para_beam_stiff}
\begin{equation}
\label{eq: para_mat_nonlocal}
    \text{Isolated Material Nonlocality:}~~[K^b]=\int_{0}^{L}\frac{Eh^3}{12} [\tilde{B}_{\mathcal{H},11}]^{T} [\tilde{B}_{\mathcal{H},11}]~\mathrm{d}x_1,~~[G^b]=\int_0^L[{B}_{{\mathcal{H},1}}]^T[{B}_{\mathcal{H},1}]~\mathrm{d}x_1
\end{equation}
\begin{equation}
\label{eq: para_geo_nonlocal}
    \text{Isolated Geometric Nonlocality:}~~[K^b]=\int_{0}^{L}\frac{Eh^3}{12} [{B}_{\mathcal{H},11}]^{T} [{B}_{\mathcal{H},11}]~\mathrm{d}x_1,~~[G^b]=\int_0^L[\tilde{B}_{{\mathcal{H},1}}]^T[\tilde{B}_{\mathcal{H},1}]~\mathrm{d}x_1
\end{equation}
\end{subequations}
Note that $[\tilde{B}_{\square}]$ are the nonlocal strain-displacement matrices employing fractional-order derivatives, and the matrices $[{B}_{\square}]$ are their local elastic analogue evaluated using integer-order derivatives as shown in \cite{reddy2014introduction}. It is clear that the stiffness terms evaluated using integer-order matrices corresponds to local elasticity. For the case of isolated material nonlocality with geometric stiffness terms being local, the softening effect introduced in $[K^b]$ by the fractional-order derivatives is expected to lead to lower values of critical load. In contrast, for the case of isolated geometric nonlocality with material stiffness terms being local, increasing the nonlocal effects would reduce the geometric stiffness, hence it will increase the critical load. These observations simply follow from the Rayleigh-Ritz expression in Eq. \eqref{eq: crit_load}. In obtaining the above results we assumed a constant length scale throughout: $l_A=l_B=l_f$.

The results for the above mentioned parametric studies, conducted for different boundary conditions, are presented in Tables~\ref{tab: cc_beam_para} and \ref{tab: ss_beam_para}. For isolated material nonlocality ($[K^b]$: Nonlocal \& $[G^b]$: Local), the critical load decreases monotonically as the degree of nonlocality increases (which is achieved either by decreasing $\alpha$ or by increasing $l_f$), irrespective of the boundary conditions. Similarly, for isolated geometric nonlocality ($[K^b]$: Local \& $[G^b]$: Nonlocal), the critical load increases with an increasing degree of nonlocality. It immediately follows that, when nonlocality is considered simultaneously in both these stiffness terms, the critical load (Table \ref{tab: beam}) is the result of the \textit{net} effect of these competing terms. This analysis explains the non-monotonic variation of the critical load with an increasing degree nonlocality. 

\begin{table}[h]
    \centering
    \begin{tabular}{c |c c c c| c c c c}
    \hline\hline
    & \multicolumn{4}{c|}{Material Nonlocality}& \multicolumn{4}{c}{Geometric Nonlocality}\\
    \hline
      $\alpha$& 1.0 & 0.9 & 0.8 & 0.7& 1.0 & 0.9 & 0.8 & 0.7\\
    \hline
       $l_f/L=$ 0.2 & 4.000 & 3.471 & 3.014 & 2.600 & 4.000 & 4.282 & 4.555 & 4.821\\
       $l_f/L=$ 0.4 & 4.000 & 2.912 & 2.124 & 1.551 & 4.000 & 4.893 & 5.930 & 7.130\\
       $l_f/L=$ 0.6 & 4.000 & 2.559 & 1.610 & 0.990 & 4.000 & 5.537 & 7.676 & 10.682\\
       $l_f/L=$ 0.8 & 4.000 & 2.411 & 1.421 & 0.810 & 4.000 & 5.986 & 9.068 & 13.999\\
       $l_f/L=$ 1.0 & 4.000 & 2.330 & 1.335 & 0.748 & 4.000 & 6.274 & 9.978 & 16.223\\
        \hline\hline
    \end{tabular}
    \caption{Non-dimensional critical loads for a double clamped (CC) beam for different values of the fractional constitutive parameters. Results are presented by artificially separating either the material or the geometric nonlocality in order to track their individual effect on the critical load.}
    \label{tab: cc_beam_para}
\end{table}
\begin{table}[b!]
    \centering
    \begin{tabular}{c |c c c c| c c c c}
    \hline\hline
    & \multicolumn{4}{c|}{Material Nonlocality}& \multicolumn{4}{c}{Geometric Nonlocality}\\
    \hline
      $\alpha$& 1.0 & 0.9 & 0.8 & 0.7& 1.0 & 0.9 & 0.8 & 0.7\\
    \hline
       $l_f/L=$ 0.2 & 1.000 & 0.978 & 0.958 & 0.939 & 1.000 & 1.095 & 1.183 & 1.263\\
       $l_f/L=$ 0.4 & 1.000 & 0.937 & 0.877 & 0.815 & 1.000 & 1.220 & 1.464 & 1.735\\
       $l_f/L=$ 0.6 & 1.000 & 0.887 & 0.776 & 0.653 & 1.000 & 1.362 & 1.841 & 2.481\\
       $l_f/L=$ 0.8 & 1.000 & 0.840 & 0.689 & 0.536 & 1.000 & 1.482 & 2.201 & 3.304\\
       $l_f/L=$ 1.0 & 1.000 & 0.803 & 0.630 & 0.469 & 1.000 & 1.568 & 2.478 & 3.986\\
        \hline\hline
    \end{tabular}
    \caption{Non-dimensional critical loads for a simply-supported (SS) beam for different values of the fractional constitutive parameters. Results are presented by artificially separating either the material or the geometric nonlocality in order to track their individual effect on the critical load.}
    \label{tab: ss_beam_para}
\end{table}

Additional observations may be drawn for the non-monotonic results in Table \ref{tab: beam} following the parametric studies in Tables \ref{tab: cc_beam_para} and \ref{tab: ss_beam_para}. As observed in previous works using fractional-order approaches \cite{patnaik2019FEM,sidhardh2020geometrically} and strain-driven integral approaches \cite{taghizadeh2016beam}, the decrease in material stiffness of a SS beam, due to nonlocal elasticity, is less pronounced when compared to a CC beam. This suggests that the reduction in the material stiffness matrix $[K^b]$ due to the fractional-order parameters (i.e. to the nonlocal effect) is offset, and subsequently dominated, by the simultaneous decrease in geometric stiffness matrix $[G^b]$ (see Eq.~\eqref{eq: beam_algebraic}).

The above observations are in contrast with the classical integer-order models for nonlocal elasticity. As discussed in the introduction, classical theories predict a consistent reduction in the critical load with increasing degree of nonlocality. This observation was drawn for both the integral\cite{zhu2017buckling,tuna2017bending,norouzzadeh2017finite} and differential\cite{taghizadeh2016beam,phadikar2010variational,narendar2011buckling} models of nonlocal elasticity. While in the case of integral-models, this effect was attributed to the reduction in material stiffness (numerator in Eq.~\eqref{eq: crit_load}), in differential models, it was caused by the the increase in geometric stiffness (denominator in Eq.~\eqref{eq: crit_load}). The current study highlights that the fractional-order continuum theory for nonlocal elasticity does not present such a monotonic decrease in critical load due to the competing effect resulting from simultaneous decrease in material and geometric stiffness terms, upon increasing the degree of nonlocality. We emphasize that the parametric studies in Tables \ref{tab: cc_beam_para} and \ref{tab: ss_beam_para} that illustrate that the non-monotonic trends in Table \ref{tab: beam} are unlike the paradoxical observations noted in literature for integer-models for nonlocal elasticity \cite{khodabakhshi2015unified,challamel2016buckling}. The paradoxical results for Eringen's integral and differential models stem from {a non positive-definite deformation energy density which leads to ill-posed and non self-adjoint governing equations} \cite{challamel2014nonconservativeness}. However, the fractional approach is based on a positive-definite deformation energy density and well-posed self-adjoint governing equations, which directly overcomes this issue. Thus, a consistent reduction of the stiffness (both material and geometric), with increasing degree of nonlocality and independently of the loading and boundary conditions\cite{patnaik2019FEM,patnaik2020plates}, is obtained. 
In addition to the above quantitative discussion, a more detailed qualitative comparison of the current approach, based on fractional-order continuum theory, with existing classical nonlocal approaches is provided in \S \ref{subsec: comparison}.

\begin{figure}[h]
    \centering
    \includegraphics[width=0.45\textwidth]{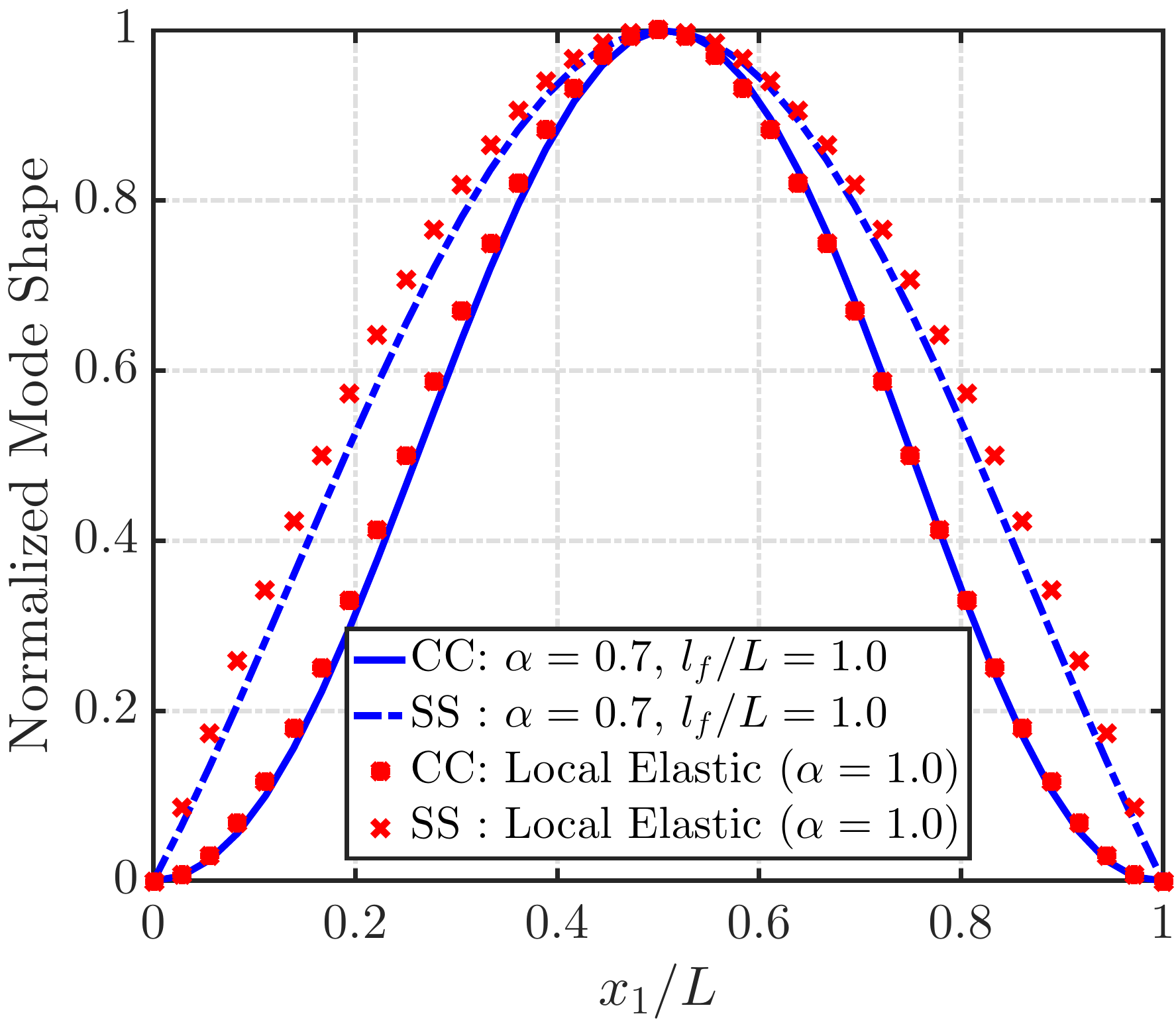}
    \caption{Comparison of the buckling mode shapes obtained from either classical or fractional-order beam theory under different boundary conditions.}
    \label{fig: mode_shape_beam}
\end{figure}

Before proceeding further, we present the transverse mode shapes corresponding to the critical load of the beam and calculated by either integer- \cite{timoshenko2009theory} or fractional-order approaches are compared in Fig. \ref{fig: mode_shape_beam}. {While slight changes are noted in the curvature of buckling mode shapes for nonlocal beams, this effect is marginal even for very a pronounced degree of nonlocality, such as for $\alpha=0.7$ and $l_f/L=1$. These} results highlight that the inclusion of the nonlocal effects via fractional-order modeling presents {minimal effects} on the buckling mode shape.

\subsection{Plates}
Consistently with the assumptions for the Kirchhoff plate theory, we selected a square plate with $a=b$ and an aspect ratio $a/h=100$ to perform the current study. The 2D f-FEM model used in this section was developed and validated in \cite{patnaik2020plates}. Both a state of uniaxial compression $N_1=N_0$ and of biaxial compression $N_2=N_1=N_0$  were investigated. The critical loads were non-dimensionalized as follows {\cite{reddy2003mechanics}}:
\begin{equation}
    \label{eq: nondim_plate}
    \overline{N}_0=N_0 \times \frac{b^2}{\pi^2 D}, ~~~~D=\frac{Eh^3}{12(1-\nu^2)}
\end{equation}
Note that the 2D f-FEM requires the evaluation of a convolution integral, along both the $x_1$ and $x_2$ directions, in order to account for the nonlocal interactions. Thus, an appropriate choice for the dynamic rates of convergence is required in both directions. For this purpose, we define $\mathcal{N}^{inf}_1=l_{f}/l_{e_1}$ and $\mathcal{N}^{inf}_2=l_{f}/l_{e_2}$ where $l_{e_1}$ and $l_{e_2}$ are the sizes of the uniform FE mesh along the $x_1$ and $x_2$ directions, respectively. The convergence of the 2D f-FEM used in this study is established for a simply supported plate (SSSS) and subject to a uniaxial compressive load along the $x_1$ direction. The results of this study for different values of the fractional-order parameters are reported in Table \ref{tab: convergence_comp_plate}. Excellent convergence of the normalized critical loads is achieved, with differences of $<$1\% between successive refinements of the mesh. Following this convergence study, we used $N^{inf}_1\times N^{inf}_2=8\times 8$ for the all the subsequent analyses.  

\begin{table}[h]
    \centering
    \begin{tabular}{c | c |c c c c}
    \hline\hline
       \multirow{2}{14em}{ Fractional Horizon Length, $l_f$} & \multirow{2}{6em}{$N^{inf}_1\times N^{inf}_2$} & \multicolumn{4}{c}{fractional-order, $\alpha$}\\
         \cline{3-6}
         &  & $\alpha=1.0$& $\alpha=0.9$& $\alpha=0.8$ & $\alpha=0.7$ \\
         \hline\hline
          \multirow{4}{5em}{$l_f=0.5a$} & $4\times 4$ & 4 & 4.120 & 4.183& 4.178\\
          & $6\times 6$ & 4 & 4.099 & 4.148 & 4.138\\
          & $8\times 8$ & 4 & 4.089 & 4.131 & 4.118\\
          & $10\times 10$ & 4 & 4.083 & 4.121 & 4.107\\
          \hline
          \multirow{4}{5em}{$l_f=a$} & $4\times 4$ & 4 & 4.258 & 4.445 & 4.545\\
          & $6\times 6$ & 4 & 4.258 & 4.452 & 4.563\\
          & $8\times 8$ & 4 & 4.259 & 4.456 & 4.573\\
          & $10\times10$ & 4 & 4.259 & 4.460 & 4.577\\
          \hline\hline
    \end{tabular}
    \caption{2D f-FEM mesh convergence study. Non-dimensional critical load of a SSSS plate for different values of the fractional parameters. In all cases, $N^{inf}=8$ guarantees a difference within 1\% between successive refinements.}
    \label{tab: convergence_comp_plate}
\end{table}

As previously mentioned, the effect of the fractional-order nonlocality on the critical load was studied for two different cases of external loading: (1) uniaxial compression along $x_1$; (2) biaxial compression with equal loads applied along $x_1$ and $x_2$. The critical loads for uniaxial and biaxial compression are presented in Table \ref{tab: plate_uni} and Table \ref{tab: plate_bi}, respectively, for different values of the fractional-order parameters $\alpha$ and $l_f$. The tabulated results correspond to two sets of boundary conditions \cite{reddy2006theory}:
\begin{equation}
\begin{split}
    \text{Simply supported (SSSS):}&~~~\begin{cases}
    x_1=0,a: & w_0=\frac{\partial w_0}{\partial x_2}=0\\
    x_2=0,b: & w_0=\frac{\partial w_0}{\partial x_1}=0
    \end{cases}\\
    \text{Clamped (CCCC):}&~~~\begin{cases}
    x_1=0,a: & w_0=\frac{\partial w_0}{\partial x_1}=\frac{\partial w_0}{\partial x_2}=0\\
    x_2=0,b: & w_0=\frac{\partial w_0}{\partial x_1}=\frac{\partial w_0}{\partial x_2}=0\\
    \end{cases}
\end{split}
\end{equation}
where $w_0$ is the generalized displacement coordinate introduced in Eq. \eqref{eq: kirch_def}. 

\begin{table}[h]
    \centering
    \begin{tabular}{c |c c c c| c c c c}
    \hline\hline
    & \multicolumn{4}{c|}{SSSS}& \multicolumn{4}{c}{CCCC}\\
    \hline
      $\alpha$& 1.0 & 0.9 & 0.8 & 0.7&  1.0 & 0.9 & 0.8 & 0.7\\
    \hline
       $l_f/a=$ 0.4& 4 & 4.109 & 4.190 & 4.233  & 10.076 & 9.874 & 9.637 & 9.334 \\
       $l_f/a=$ 0.6& 4 & 4.146 & 4.234 & 4.247  & 10.076 & 9.834 & 9.544 & 9.162  \\
       $l_f/a=$ 0.8& 4 & 4.238 & 4.417 & 4.517  & 10.076 & 10.017 & 10.005 & 10.023 \\
       $l_f/a=$ 1.0& 4 & 4.259 & 4.456 & 4.573  & 10.076 & 9.981 & 9.950 & 9.965  \\
        \hline\hline
    \end{tabular}
    \caption{Non-dimensional critical loads for a plate subject to uniaxial compression and various boundary conditions. Results are presented for different values of the fractional constitutive parameters.}
    \label{tab: plate_uni}
\end{table}

\begin{table}[b!]
    \centering
    \begin{tabular}{c |c c c c| c c c c}
    \hline\hline
    & \multicolumn{4}{c|}{SSSS}& \multicolumn{4}{c}{CCCC}\\
    \hline
      $\alpha$& 1.0 & 0.9 & 0.8 & 0.7& 1.0 & 0.9 & 0.8 & 0.7\\
    \hline
       $l_f/a=$ 0.4& 2 & 2.055 & 2.098 & 2.125 & 5.304 & 5.112 & 4.917 & 4.706 \\
       $l_f/a=$ 0.6& 2 & 2.074 & 2.120 & 2.136 & 5.304 & 5.086 & 4.862 & 4.613  \\
       $l_f/a=$ 0.8& 2 & 2.119 & 2.210 & 2.266 & 5.304 & 5.212 & 5.146 & 5.099 \\
       $l_f/a=$ 1.0& 2 & 2.129 & 2.230 & 2.293 & 5.304 & 5.196 & 5.123 & 5.075  \\
        \hline\hline
    \end{tabular}
    \caption{Non-dimensional critical loads for a plate subject to biaxial compression and various boundary conditions. Results are presented for different values of the fractional constitutive parameters.}
    \label{tab: plate_bi}
\end{table}

As evident from the Tables \ref{tab: plate_uni},\ref{tab: plate_bi}, the critical loads of fractional-order plates show a non-monotonic variation with increasing degree of nonlocality. While this observation deviates from the conclusions of studies based on classical integer-order nonlocal theories, it does agree with the results reported in \S\ref{ssec: beam_results} for nonlocal beams. Similar to the discussion in \S\ref{ssec: beam_results}, this difference is a direct result of the simultaneous effect of the nonlocal response on both the geometric and material stiffness of the fractional-order plate. The increasing degree of nonlocality (achieved, for example, by reducing $\alpha$ or by increasing $l_f$) results in a consistent reduction of the system stiffness matrices $[K^p]$ and $[G^p]$ given in Eq. \eqref{eq: stiff_plate}. However, the individual reductions in these stiffness terms have a competing effect on the critical buckling load of the nonlocal structure, which is evident from the Rayleigh-Ritz given in Eq. \eqref{eq: crit_load}. To better illustrate the contrasting effects of the decreasing material and geometric stiffness terms on the critical load, we perform a study similar to that conducted for beams. In an analogous way, we suppress the effect of nonlocality on one of the stiffness terms (either material \textit{or} geometric) and evaluate the corresponding critical load. This approach allows isolating the nonlocal effect of each individual stiffness term on the critical load. The suppression of the nonlocal effect is achieved by replacing the fractional-order derivatives with their integer-order counterparts. These system stiffness matrices can be obtained analogous to Eq. \eqref{eq: para_beam_stiff}. The results of this parametric study conducted for a uniaxial compressive load, different boundary conditions, and different levels of nonlocality are presented in Tables \ref{tab: cc_plate_para} and \ref{tab: ss_plate_para}. 
%
%
In these analyses, similar to the results obtained for beams, we assumed isotropic length scales in both the directions at all points within the domain of the plate. Comparing the numerical results, it is clear that in the case of isolated material nonlocality (material stiffness: nonlocal; geometric stiffness: local) a monotonic reduction of the critical load is associated with an increasing degree of nonlocality. However, for the case of isolated geometric nonlocality (material stiffness: local; geometric stiffness: nonlocal) the situation is inverted. These competing effects result in the non-monotonic variation of the critical load in fractional-order nonlocal plates with increasing degree of nonlocality. 

\begin{table}[h]
    \centering
    \begin{tabular}{c |c c c c| c c c c}
    \hline\hline
    & \multicolumn{4}{c|}{Material Nonlocality}& \multicolumn{4}{c}{Geometric Nonlocality}\\
    \hline
      $\alpha$& 1.0 & 0.9 & 0.8 & 0.7& 1.0 & 0.9 & 0.8 & 0.7\\
    \hline
       $l_f/L=$ 0.4 & 5.304 & 3.763 & 2.650 & 1.845 & 5.304 & 6.880 & 8.909 & 11.538\\
       $l_f/L=$ 0.6 & 5.304 & 3.437 & 2.197 & 1.378 & 5.304 & 7.583 & 10.901 & 15.832\\
       $l_f/L=$ 0.8 & 5.304 & 3.265 & 1.987 & 1.188 & 5.304 & 8.098 & 12.496 & 19.635\\
       $l_f/L=$ 1.0 & 5.304 & 3.123 & 1.817 & 1.039 & 5.304 & 8.467 & 13.663 & 22.448\\
        \hline\hline
    \end{tabular}
    \caption{Non-dimensional critical loads for a fully clamped (CCCC) plate subject to biaxial compression for different values of fractional constitutive parameters. Results are presented {by artificially separating} either the material or the geometric nonlocality {in order to track their individual effect on the critical load.}}
    \label{tab: cc_plate_para}
\end{table}

\begin{table}[b!]
    \centering
    \begin{tabular}{c |c c c c| c c c c}
    \hline\hline
    & \multicolumn{4}{c|}{Material Nonlocality}& \multicolumn{4}{c}{Geometric Nonlocality}\\
    \hline
      $\alpha$& 1.0 & 0.9 & 0.8 & 0.7& 1.0 & 0.9 & 0.8 & 0.7\\
    \hline
       $l_f/L=$ 0.4 & 2.000 & 1.732 & 1.512 & 1.310 & 2.000 & 2.561 & 3.246 & 4.089\\
       $l_f/L=$ 0.6 & 2.000 & 1.593 & 1.274 & 0.999 & 2.000 & 2.826 & 3.981 & 5.618\\
       $l_f/L=$ 0.8 & 2.000 & 1.496 & 1.119 & 0.821 & 2.000 & 3.055 & 4.694 & 7.317\\
       $l_f/L=$ 1.0 & 2.000 & 1.430 & 1.024 & 0.718 & 2.000 & 3.195 & 5.132 & 8.365\\
        \hline\hline
    \end{tabular}
    \caption{Non-dimensional critical loads for a simply supported (SSSS) plate subject to biaxial compression for different values of fractional constitutive parameters. Results are presented {by artificially separating} either the material or the geometric nonlocality {in order to track their individual effect on the critical load.}}
    \label{tab: ss_plate_para}
\end{table}

Further, we note from the Table \ref{tab: plate_uni} that the critical load for the nonlocal CCCC plate is lower than its local elastic analogue, while for the nonlocal SSSS plate the critical load is higher compared to the local elastic case. {These contrasting observations can be explained by considering the stronger effect of nonlocal interactions on decreasing the material stiffness for plates subject to stiffer boundary conditions \cite{patnaik2020plates,patnaik2020geometrically}. More specifically, a stronger reduction is noted in the material stiffness of plates subject to clamped boundary conditions when compared to simply supported boundary conditions. As discussed previously in the case of fractional-order SS beams, the weak reduction in material stiffness for SSSS plates is further dominated by the simultaneous decrease in geometric stiffness. Thus, the net result of nonlocal interactions on material and geometric stiffness terms is an increase in the critical load for SSSS plates. In contrast to this, the marked decrease in material stiffness for the clamped plate ensures lower critical load for CCCC fractional-order plates.}



\begin{figure}[b!]
    \centering
    \includegraphics[width=0.45\textwidth]{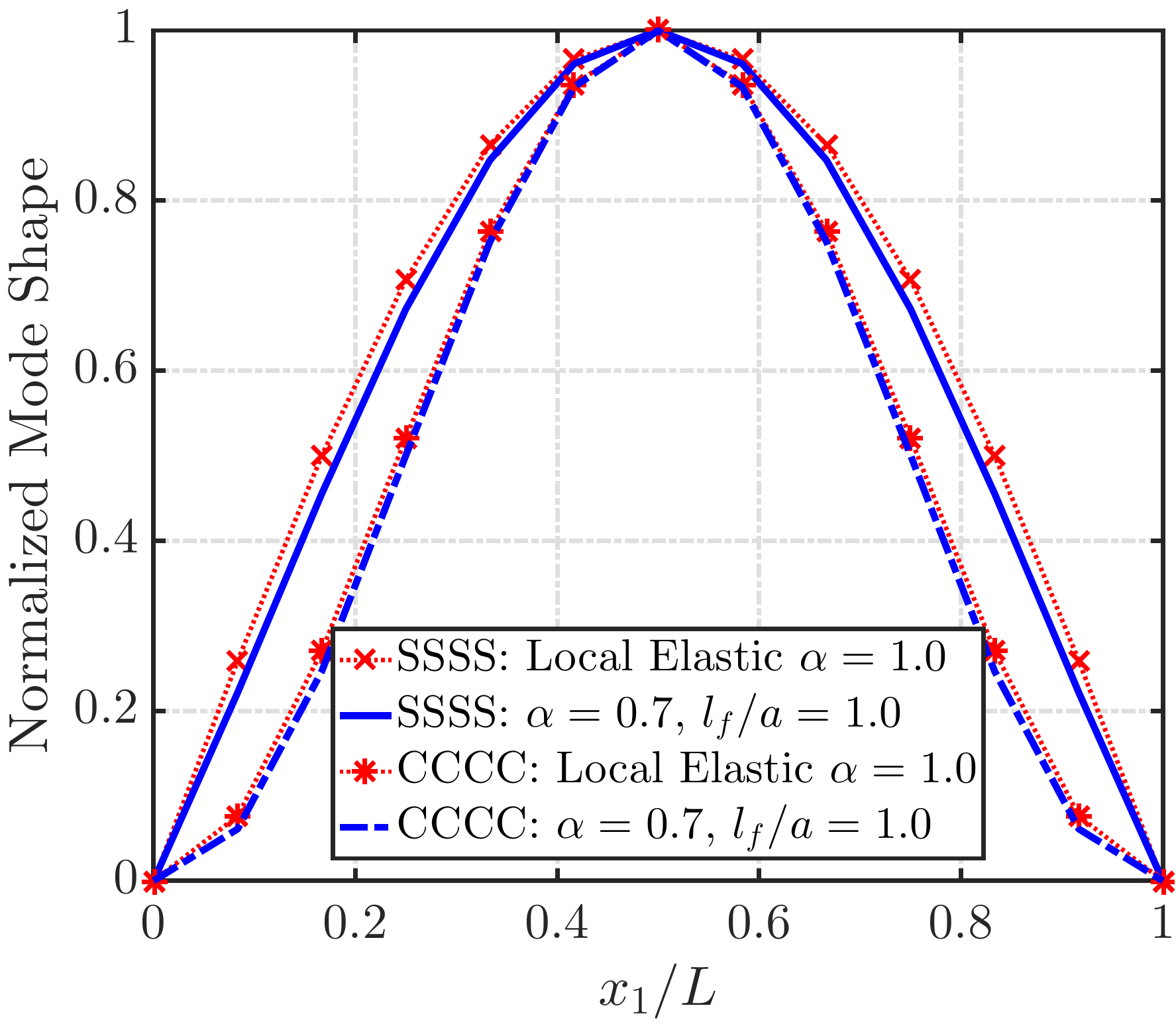}
    \caption{{Comparison of the buckling mode shapes for biaxial compression obtained from either classical or fractional-order plate theory under different boundary conditions.}}
    \label{fig: mode_shape_plate}
\end{figure}

Finally, we compare the transverse mode shape along the length ($x_1$) at $x_2=b/2$ corresponding to the critical load of the beams modeled via both integer-order \cite{timoshenko2009theory} and fractional-order approaches. This comparison is illustrated in Fig. \ref{fig: mode_shape_plate}. {As seen in our previous study over fractional-order beams in Fig. \ref{fig: mode_shape_beam}, marginal effects of the nonlocal interactions are realized on the mode shape of plate corresponding to critical buckling load. This is substantiated by the weak changes in curvature of the normalized modes for critical buckling of the nonlocal plate when compared with local analogues even for a case with high degree of nonlocality ($\alpha=0.7$ and $l_f/a=1.0$).}

\subsection{Comparison with existing integer-order nonlocal theories}
\label{subsec: comparison}
In this section, we compare the effects of the nonlocal interactions on the critical load when accounted for by either fractional-order or integer-order theories of nonlocal elasticity. For this purpose, we begin with the two-phase (i.e. local/nonlocal) integral model proposed in \cite{altan1989uniqueness}. {This choice is motivated by several studies in the literature that are based on this theory \cite{zhu2017buckling} or on simplified models derived from it \cite{tuna2016exact,tuna2017bending,taghizadeh2016beam,norouzzadeh2017pre} to study nonlocal effects on the critical load.} The nonlocal constitutive relations take the form of a Fredholm equation of the second kind:
\begin{equation}
    \label{eq: ering_contt}
    \tilde{\bm{\sigma}}(\textbf{X})=\int_{\Omega}~\overline{\mathcal{A}}(\textbf{X},\bm{\xi})~\textbf{C}:\bm{\epsilon}(\bm{\xi})~\mathrm{d}\bm{\xi}, ~~~\overline{\mathcal{A}}=\chi_1 \delta(\textbf{X},\bm{\xi})+\chi_2 \mathcal{A}(\textbf{X},\bm{\xi})
\end{equation}
where $\bm{\epsilon}$ is the local strain evaluated using the classical integer-order strain-displacement relations. Also, $\delta(\textbf{X},\bm{\xi})$ is the Dirac-delta defined at $\textbf{X}$, $\overline{\mathcal{A}}(\textbf{X},\bm{\xi})$ is the attenuation function, $\chi_1$ and $\chi_2$ are positive material constants that satisfy: $\chi_1+\chi_2=1$. 
Eringen's integral nonlocal model, corresponding to a Fredholm equation of the first kind, can be obtained for appropriate choices of $\chi_1$ and $\chi_2$\cite{polizzotto2001nonlocal}. A detailed discussion on the comparison of the constitutive laws for the fractional-order and integer-order models of \textit{linear} nonlocal elasticity is provided in \cite{patnaik2019FEM}. In \cite{patnaik2019FEM}, the linear fractional-order continuum theory was obtained from the integral Eringen model assuming suitable choices of the attenuation function and the domain of influence. However, this equivalence holds only for the linear kinematic relations. For the geometrically nonlinear models of nonlocal elasticity, the constitutive models that follow the fractional-order kinematic relations in Eq.~\eqref{eq: finite_fractional_strain} cannot be deduced from Eringen's model in an analogous manner. This observation is relevant to understand the differences in critical load observed when using the fractional-order and the classical integer-order theories of nonlocal elasticity \cite{taghizadeh2016beam,zhu2017buckling,tuna2017bending,norouzzadeh2017pre}. 

To elaborate further, we focus on the case of the nonlocal Euler-Bernoulli beam. The weak statement for the transverse equilibrium equation for the fractional-order nonlocal beam, that follows from Eq.~\eqref{eq: weak_statement_beam}, is given by:
\begin{equation}
    \int_0^L\delta\{\Delta^b\}^T\left([K^b]-{N}_{0}[G^b]\right)\{\Delta^b\}~\mathrm{d}x_1=\{0\}
\end{equation}
where
\begin{equation}
\label{eq: stiff_frac}
    [K^b]=\int_{0}^{L}\frac{Eh^3}{12} \left(D_{x_1}^{\alpha}w_0(x_1)\right)^2~\mathrm{d}x_1,~~~[G^b]=\int_0^L\left(D_{x_1}^{\alpha}w_0(x_1)\right)^2~\mathrm{d}x_1
\end{equation}
$[K^b]$ and $[G^b]$ are the material and geometric stiffness matrices. Similarly, the expressions for these matrices derived following the integer-order nonlocal constitutive relations given in Eq.~\eqref{eq: ering_contt} would be\cite{taghizadeh2016beam,norouzzadeh2017pre}:
\begin{equation}
\label{eq: stiff_erin}
    [K^b]=\int_{0}^{L}\int_{0}^{L}\frac{Eh^3}{12} \mathcal{A}(x_1,x_1')\left(D_{x_1}^{2}w_0(x_1)\right)\left(D_{x_1}^{2}w_0(x_1')\right)~\mathrm{d}x_1\mathrm{d}x_1',~~~[G^b]=\int_0^L\left(D_{x_1}^{1}w_0\right)^2~\mathrm{d}x_1
\end{equation}
Appropriate choices for the material constants $\chi_1$ and $\chi_2$ can reduce this model to the single-phase Eringen's model\cite{tuna2017bending}. For the sake of comparison, we provide below the expressions for the stiffness matrices evaluated assuming local elasticity \cite{reddy2014introduction}:
\begin{equation}
\label{eq: stiff_local}
    [K^b]=\int_{0}^{L}\frac{Eh^3}{12} \left(D_{x_1}^{2}w_0(x_1)\right)^2~\mathrm{d}x_1,~~~[G^b]=\int_0^L\left(D_{x_1}^{1}w_0\right)^2~\mathrm{d}x_1
\end{equation}
We note that the material and geometric stiffness matrices for the fractional-order theory of nonlocal elasticity {given in Eq.~\eqref{eq: stiff_frac}} include fractional derivatives. The fractional derivatives within the definition of these stiffness terms capture the nonlocal interactions across the domain. An increase in the degree of nonlocality (either by reducing the fractional-order $\alpha$ or by increasing the length scale $l_f$) reduces the numerical value of the fractional-order derivatives, and hence of the corresponding stiffness term. This softening effect is evident from the results of the parametric studies in Tables \ref{tab: cc_beam_para}, \ref{tab: ss_beam_para}, \ref{tab: cc_plate_para} and \ref{tab: ss_plate_para}. In the case of Eringen's integral models, the stiffness matrix $[K^b]$ in Eq.~\eqref{eq: stiff_erin} includes nonlocal interactions across the domain and undergoes reduction with increasing degree of nonlocality. However, the geometric stiffness $[G^b]$ in this equation, remains identical to its local form in Eq. \eqref{eq: stiff_local}, as also observed in \cite{tuna2017bending}. 
{The decreasing values of stiffness $[K^b]$ with increasing degree of nonlocality, while $[G^b]$ remains constant, explains the lower critical load for nonlocal solids predicted by Eringen's theory\cite{tuna2017bending,zhu2017buckling}. This observation is akin to our parametric studies on fractional-order beams and plates over material nonlocality (see Eq. \eqref{eq: para_mat_nonlocal}).}
{Therefore, we conclude that fractional-order continuum theories differ from the Eringen's integral nonlocal models primarily in their formulation of the geometric stiffness. Unlike the integer-order theory, the geometric stiffness matrix for the fractional-order solid is also modified by the presence of fractional-order derivatives. 
The corresponding reduction in geometric stiffness $[G^b]$ due to nonlocal interactions is evident from the parametric studies conducted from Eq. \eqref{eq: para_geo_nonlocal}. The simultaneous reduction in both material and geometric stiffness terms has competing effects on the critical load. 
}

For the sake of completeness, we also compare the fractional-order model with the Eringen's differential model of nonlocal elasticity. The nonlocal stiffness terms evaluated for a Euler-Bernoulli beam using the differential model of nonlocal elasticity are\cite{pradhan2012nonlocal}:
\begin{equation}
\label{eq: stiff_diff}
    [K^b]=\int_{0}^{L}\frac{Eh^3}{12} \left(D_{x_1}^{2}w_0(x_1)\right)^2~\mathrm{d}x_1,~~~[G^b]=\int_0^L\left[ l_e^2\left(D_{x_1}^{2}w_0(x_1)\right)^2+\left(D_{x_1}^{1}w_0\right)^2\right]~\mathrm{d}x_1
\end{equation}
where $l_e$ is the characteristic length scale \cite{eringen1983differential}. Comparing the above expressions with Eq. \eqref{eq: stiff_local}, it is clear that the differential model predicts a modification in the geometric stiffness caused by the nonlocal interactions. However, the material stiffness is identical to the classical local elasticity case given in Eq. \eqref{eq: stiff_local}. Therefore, the nonlocal effects on the critical load are realized only by a modification of the geometric stiffness matrix. {This observation is also in contrast to the effect of nonlocal elasticity being realized on both the stiffness matrices following fractional-order continuum theories (see Eq. \eqref{eq: stiff_frac}). From Eq. \eqref{eq: stiff_diff}, we note increasing values of stiffness $[G^b]$ with increasing degree of nonlocality, while $[K^b]$ remains constant. This explains the lower critical load for nonlocal solids predicted by Eringen's differential models\cite{norouzzadeh2017pre,pradhan2009small}. Note that similar observation was noted above from our parametric study on fractional-order beams and plates for isolated case of geometric nonlocality (see Eq. \eqref{eq: para_geo_nonlocal}).}

\section{Conclusions}
The present study extends the fractional-order continuum theory framework to perform stability analysis of nonlocal solids. Thanks to the thermodynamically consistent and positive-definite form of the deformation energy density afforded by the fractional-order formulation, we can apply energy methods to perform the stability analysis. {We reiterate that the geometrically nonlinear models for nonlocal elasticity, available within the framework of fractional calculus, allow the stability analysis to be conducted for nonlocal solids. As part of this approach, a general stability analysis is carried out for fractional-order solids employing the Lagrange-Dirichlet theorem. This allows studying scale effects, nonlocality, and heterogeneity on the stability of complex solids and interfaces. Specializing it to the case of linear buckling, we derive the Rayleigh-Ritz expression for the critical load of nonlocal solids. Our results show the capability of fractional-order models to realize nonlocal effects on the material and geometric stiffness terms. This is unlike the classical integral and differential models for nonlocal elasticity, where the nonlocal effects are restricted to only one of the stiffness terms. Thus, a more accurate account of the effects of nonlocal interactions on the stability of structures is realized by using fractional-order theories. To illustrate this observation in a quantitative manner, the stability analysis was performed for fractional-order beams and plates, and the resulting eigenvalue problems were solved using a finite-element numerical method developed for fractional-order boundary value problems. 
}

\section*{Acknowledgements}
The following work was supported by the Defense Advanced Research Project Agency (DARPA) under the grant \#D19AP00052, and the National Science Foundation (NSF) under the grant DCSD \#1825837. The content and information presented in this manuscript do not necessarily reflect the position or the policy of the government. The material is approved for public release; distribution is unlimited.

\bibliographystyle{unsrt}
\bibliography{report}

\end{document}